\documentclass[sigconf]{acmart}
\AtBeginDocument{%
  }


\usepackage[many]{tcolorbox}
\usepackage{amsthm}
\newtheorem{definition}{Definition}
\usepackage{url}
\usepackage{hyperref}
\usepackage{nicefrac}
\usepackage{siunitx}
\usepackage{array,framed}
\usepackage{pdfpages}
\usepackage{booktabs}
\usepackage[normalem]{ulem}
\usepackage{
  color,
  float,
  epsfig,
  graphics,
  graphicx,
  subcaption
}

\usepackage{amsfonts}
\usepackage{latexsym,fancyhdr}
\usepackage{enumerate}
\usepackage{algorithm,algorithmic}
\usepackage{xparse} 
\usepackage{xspace}
\usepackage{multirow}
\usepackage{csvsimple}
\usepackage{enumitem}
\usepackage{bbding}
\usepackage{pifont}
\usepackage{colortbl}
\usepackage{threeparttable}
\usepackage{longtable}
\usepackage{soul}

\usepackage{
  tikz,
  pgfplots,
  pgfplotstable
}

\usetikzlibrary{
  shapes.geometric,
  arrows,
  external,
  pgfplots.groupplots,
  matrix
}

\usepackage{mathtools,}

\newcommand{\llamaseven}{\textsc{Llama-2-7b-chat}\xspace}
\newcommand{\llamaseventy}{\textsc{Llama-2-70b-chat}\xspace}

\newcommand{\gemmanine}{\textsc{gemma-2-9b-it}\xspace}
\newcommand{\llamathirteen}{\textsc{Llama-2-13b-chat}\xspace}
\newcommand{\qwenseven}{\textsc{Qwen2.5-7b-Instruct}\xspace}
\newcommand{\qweneight}{\textsc{Qwen3-8b}\xspace}

\newcommand{\llamathreeit}{\textsc{Llama-3.1-8b-Instruct}\xspace}
\newcommand{\tool}{\textsc{DeCNIP}\xspace}

\newcommand{\llamaguard}{\textsc{Llama-Guard-3}\xspace}
\newcommand{\jailbench}{\texttt{JailbreakBench}\xspace}

\definecolor{lightgray}{gray}{0.9} 
\definecolor{BestColor}{rgb}{0.8, 0.0, 0.0} 
\definecolor{SecondBestColor}{rgb}{0.0, 0.0, 0.7} 

\newcommand{\best}[1]{\textbf{\textcolor{BestColor}{#1}}}
\newcommand{\secondbest}[1]{\underline{\textcolor{SecondBestColor}{#1}}}

\definecolor{lightgray}{gray}{0.9} 

\definecolor{WarningColor}{rgb}{0.6, 0.2, 0.0} 

\newcommand{\worst}[1]{\textbf{\textcolor{WarningColor}{#1}}}
\newcommand{\secondworst}[1]{\underline{\textcolor{WarningColor}{#1}}}

\begin{abstract}
Large language models (LLMs) have advanced rapidly across domains, yet their growing complexity increases vulnerability to security threats such as \emph{backdoor attacks}, where hidden triggers induce malicious or unintended outputs.
Existing defenses generally fall into inference-time detection or training-time mitigation, yet they face two fundamental limitations. First, they are primarily designed for fine-tuning-based backdoors, particularly those embedded in PEFT modules, and therefore fail to address more insidious model-editing attacks that bypass conventional training pipelines. Second, they are typically developed around simple classification settings and do not naturally extend to the open-ended generation characteristics of LLMs. Consequently, these methods focus on surface-level behavioral patterns while neglecting the deeper representational causes of malicious activations. This lack of mechanistic understanding forces defenses to depend on empirical heuristics, limiting their robustness, generality, and practical applicability in real-world LLM deployment.

To bridge this gap,
we introduce \tool (\textbf{De}fense with \textbf{C}ritical \textbf{N}euron \textbf{I}solation \textbf{P}runing). It leverages representational analysis to identify and neutralize backdoors within a single detection and mitigation pipeline. Specifically, \tool identifies trigger-like behaviors by optimizing a cross-entropy-based loss between harmful prompts with candidate tokens and benign inputs. This deep representational discovery enables the framework to expose latent threats by uncovering the fundamental mechanisms through which triggers hijack model weights. It then isolates \textbf{Backdoor Critical Neurons (BCNs)} and prunes them selectively to remove malicious influence while preserving model utility. 
Extensive evaluations on six open-source LLMs and two benchmark datasets demonstrate that DeCNIP achieves \textbf{more than 95\% relative reduction in Attack Success Rate (ASR)}, outperforming seven state-of-the-art defenses with \textbf{only 0.1\% of the neurons intervened}. Moreover, it maintains \textbf{an average of 97\% of the model's foundational performance} on normal benchmarks, illustrating its efficacy, robustness, and scalability in securing large-scale generative models.

\end{abstract}

\begin{document}
%
\title{Defense Against LLM Backdoors using Critical Neuron Isolation Pruning}


\author{Yuxi Li}
\affiliation{%
  \institution{Huazhong University of Science and Technology}
  \country{China}
}
\email{yuxili@hust.edu.cn}

\author{Zhibo Zhang}
\affiliation{%
  \institution{Huazhong University of Science and Technology}
  \country{China}
}
\email{zhangzhibom@hust.edu.cn}

\author{Kailong Wang}
\authornote{Corresponding author.}
\affiliation{%
  \institution{Huazhong University of Science and Technology}
  \country{China}
}
\email{wangkl@hust.edu.cn}

\author{Xingshuo Han}
\affiliation{%
  \institution{Nanjing University of Aeronautics and Astronautics}
  \country{China}
}
\email{xingshuo.han@nuaa.edu.cn}

\author{Ling Shi}
\affiliation{%
  \institution{Nanyang Technological University}
  \country{Singapore}
}
\email{ling.shi@ntu.edu.sg}

\author{Haoyu Wang}
\affiliation{%
  \institution{Huazhong University of Science and Technology}
  \country{China}
}
\email{haoyuwang@hust.edu.cn}

\maketitle



%


\section{Introduction}
\label{sec:intro}

Large language models (LLMs) have advanced rapidly in recent years and now play a critical role across a wide range of domains, including industry, education, and healthcare~\cite{vajjala2025opportunitieschallengesllmseducation, säuberli2025llmspsychometricallyplausibleresponses, zhu2025medicalosllmagentbased, bian2025benchmarkingethicalsafetyrisks}. 
Nevertheless, these models face severe security threats from backdoor attacks~\cite{li2024badedit, yan-etal-2024-backdooring, hubinger2024sleeperagentstrainingdeceptive}, which involve the implantation of hidden malicious logic that remains dormant until activated by specific input triggers. Backdoor techniques targeting LLMs primarily fall into two categories: fine-tuning-based attacks, where attackers inject poisoned samples into the training pipeline, and model-editing attacks, which involve direct manipulation of model weights to embed trigger-response pairs. These attacks severely undermine model integrity, potentially leading to unauthorized data exfiltration or the generation of harmful content.

Existing defenses against backdoor attacks can be broadly divided into \emph{inference-time defense} and \emph{training-time defense}~\cite{li2022backdoorlearningsurvey, BaiBackdoorAttack}. Inference-time detection aims to identify and suppress abnormal model behaviors during inference by analyzing the relationship between inputs and outputs~\cite{sun2025peftguard, shen2025bait}. However, this class of defenses faces two major limitations. First, many existing methods are specifically tailored to identify backdoor artifacts within PEFT modules and LoRA adapters. Consequently, they often fail to generalize to attacks via model editing, which bypasses traditional fine-tuning pipelines by directly manipulating model weights, thereby significantly narrowing the practical detection scope. Second, most detection mechanisms focus exclusively on the surface-level behaviors of the model, neglecting the deeper representational causes that drive such malicious activations. 

Compared to inference-time defenses, training-time defenses attempt to neutralize backdoors before deployment by adjusting models' parameters, often through fine-tuning or model pruning. Fine-tuning-based defenses generally fail to completely remove backdoor behaviors, leaving residual vulnerabilities that attackers may still exploit. Meanwhile, pruning-based defenses effectively address the root causes of backdoors but still encounter two fundamental challenges in the context of generative LLMs. On one hand, although localizing backdoor neurons and pruning have been explored for backdoor removal in conventional deep learning models~\cite{wu2021adversarial, li2021neural, liu2018fine}, these methods are primarily designed for discriminative tasks with a closed-set label space (e.g., image or text classification). Directly transferring them to decoder-only LLMs is difficult because the nature of backdoor triggers differs significantly between classification tasks and generative language modeling. 


On the other hand, while recent pruning strategies have been proposed for language models~\cite{zhao2024defense, Santosh2025pruning}, they predominantly target encoder-only architectures (e.g., BERT, RoBERTa) where the defense objective is to rectify a flipped classification label. In contrast, backdoors in generative LLMs hijack the entire autoregressive trajectory, making the identification of ``malicious neurons'' significantly more complex and computationally expensive. Compounding this challenge is the fact that prior pruning research for LLMs has been largely optimized for inference acceleration~\cite{zhu2024dppapruningmethodlarge, sunsimple} rather than robustness. Consequently, these methods often disregard the safety-critical neurons that govern malicious activations. Applying them directly thus causes a severe drop in model utility and stability, as they fail to preserve the balance between security and the model's inherent reasoning capabilities. These limitations motivate a key question: \textit{\uline{Can we design an approach that effectively and surgically removes backdoors in decoder-only generative LLMs while preserving the model’s original performance and complex reasoning capabilities?}}

To establish a principled foundation for defense, we conduct a systematic analysis to characterize the operational mechanisms of backdoors during inference. By evaluating the model’s response to varied trigger configurations and analyzing internal hidden state evolutions across benign and harmful contexts, we ensure a rigorous assessment of how malicious logic is activated. Our investigation reveals two pivotal insights: first, we observe that the trigger's influence on model behavior is largely invariant to its specific surface form or spatial positioning within a prompt, with malicious behavior being consistently elicited across diverse token variations and locations; second, layer-wise activation analysis demonstrates that harmful prompts containing these triggers generate internal representations that are deceptively similar to those of benign queries. \textit{\uline{These findings collectively suggest that generative backdoors are not mere surface-level mappings but are encoded in deep representational structures that hijack the model’s reasoning trajectory by mimicking benign processing patterns.}}


Motivated by the insight that triggers function as mechanistic ``entry points'' within the representational space, we propose \tool, a unified framework designed to interpret and neutralize backdoors through deep activation analysis. The framework utilizes a cross-entropy-based optimization objective that compares augmented harmful prompts against benign counterparts, which enables the discovery of latent triggers by exposing the fundamental mechanisms through which they hijack model weights. By isolating \textbf{Backdoor Critical Neurons (BCNs)}, a specialized subset of neurons functionally coupled with backdoor activations, \tool selectively prunes these components to eliminate malicious influence while ensuring the model's foundational utility remains intact. Extensive evaluations involving six open-source LLMs demonstrate that our approach achieves\textbf{ over 95\% relative reduction in ASR}, which significantly outperforms seven state-of-the-art defenses while maintaining an\textbf{ average of 97\% of normal functionality} on MT-Bench, HumanEval and AlpacaGPT-52K with \textbf{only 0.1\% of the neurons intervened}, indicating the reasonability and the real-world availability of \tool. This evaluation result indicates that \tool provides a useful scenario for defending against backdoor attacks, therefore giving a possible solution for the development and construction of the LLM community.


\noindent \textbf{Contributions.} The key contributions are as follows:\vspace{-0.3cm}
\begin{itemize}[leftmargin=*]
    \item We characterize backdoor activation in decoder-only LLMs through a comprehensive analysis of triggers, demonstrating that direct migration of existing defenses fails to preserve model utility due to a neglect of internal representational dynamics.

    \item We develop \tool, a unified framework for the precise localization and pruning of BCNs, demonstrating that neutralizing backdoors requires only 0.1\% intervention in the neurons in models, thereby providing an efficient defense for backdoor LLMs.


    \item \tool outperforms seven baseline defenses on four attacks across six models, while maintaining a strong score on normal benchmarks like MT-bench, demonstrating scalability and robustness to large-scale generative architectures.
\end{itemize}
\section{Background}
\label{sec:background}

\subsection{LLM Running Process}
The vast majority of modern generative LLMs are built upon the decoder-only Transformer architecture. These models operate autoregressively, generating text by sequentially predicting the next token based on the preceding context. Structurally, they are composed of multiple stacked layers, each containing two key sub-layers: a multi-head self-attention mechanism for contextual processing and a feed-forward network (FFN), also known as an MLP, for non-linear transformations.



\noindent\textbf{Self-Attention Blocks.}  
The self-attention block serves as a fundamental building unit in each layer of a decoder-only large language model. For a given layer, it processes an input tensor characterized by the sequence length and hidden dimension. Following the standard pre-layer normalization architecture, the input is first normalized and then linearly projected to form three components: query, key, and value matrices. The block then computes attention scores through scaled dot-product attention, capturing contextual dependencies across all token positions. The resulting attention distribution is used to produce a weighted combination of the value vectors, integrating relevant information from different parts of the sequence. Finally, this attention output is combined with the original input through a residual connection, yielding an intermediate representation that serves as the input to the subsequent feed-forward network.

\noindent\textbf{Gated MLP Blocks.} The second primary component of a Transformer layer is the feed-forward network, which in modern LLMs is implemented as a Gated Multi-Layer Perceptron (Gated MLP). This block processes the intermediate representation $x_l^{\text{mid}}$. Similar to the attention block, the input first undergoes layer normalization to produce $x_l^{\text{mid-norm}}$. This normalized tensor is then passed through two parallel linear projections: an up-projection layer with weight $W_l^{\text{in}}$ and a gate layer with weight $W_l^{\text{gate}}$. The gating mechanism combines these two outputs via an element-wise product, where the gate's output is first passed through a non-linear activation function ($\sigma$, e.g., SiLU). The result is subsequently projected back to the model's hidden dimension by a down-projection layer with weight $W_l^{\text{out}}$. This is expressed as:
\begin{gather}
    x_l^{\text{ffn-norm}} = \text{LayerNorm}(x_l^{\text{ffn-in}}) \\
    x_l^{\text{ffn-out}} = \left( \sigma(x_l^{\text{ffn-norm}} W_l^{\text{gate}}) \odot (x_l^{\text{ffn-norm}} W_l^{\text{in}}) \right) W_l^{\text{out}}
\end{gather}

\subsection{LLM Backdoor Attacks}
Backdoor attacks on LLMs represent a significant security threat, where an adversary aims to implant hidden, malicious behaviors into a seemingly benign model. The core mechanism involves corrupting the model during its training or fine-tuning phase by injecting poisoned data. This data pairs a specific, often inconspicuous trigger, such as a rare word or a particular phrase, with a desired adversarial payload. Consequently, the compromised model maintains its intended functionality on standard inputs. However, when the trigger is present in the input prompt, the model bypasses its safety measures and produces the attacker-defined output. Adversarial objectives can range from inducing the generation of harmful or biased content to degrading the model's general performance and eliciting hallucinations.

Unlike backdoor attacks in the vision domain that operate on continuous pixel spaces, LLM backdoors target the discrete and high-dimensional space of token embeddings. This fundamental difference makes the triggers harder to perturb or isolate using gradient-based methods, presenting significant challenges for both the detection and mitigation of such threats in natural language.

\section{Motivation}
\label{sec:empirical}

\subsection{Understanding Backdoor Behaviors in LLMs}
\label{sec:study_method}

To gain a deeper understanding of the behavioral characteristics of backdoored models, we perform an analysis focusing on the functionality of trigger words during inference. Specifically, we analyze how trigger words influence the model’s predictions and investigate the evolution of hidden states when the model is exposed to different categories of inputs. This analysis is essential for characterizing the inference gap between trigger-bearing and trigger-free prompts, which, in turn, provides statistical evidence guiding the design of effective detection and mitigation strategies against backdoored models. Our study consists of two main components:

\noindent\textbf{Influence of Trigger Words.} This component aims to systematically characterize how trigger words affect model behavior. We examine the model’s responses when a trigger word is placed at different positions within the input (e.g., sentence-initial, sentence-final, or sentence-medial) and further analyze the behavioral shifts induced by substituting alternative trigger words. Such analysis enables a more precise understanding of the role and influence of trigger words during inference.

\noindent\textbf{Characteristics of Hidden States in Backdoor Models.} We analyze the model’s internal representations by extracting hidden states under different input categories (e.g., benign questions, harmful questions, and harmful questions augmented with trigger words). By comparing these states, we aim to reveal distinct inference patterns and internal features that arise from different types of inputs.

\subsection{Observing Model Behaviors}
\label{sec:study_imple}

\noindent\textbf{Dataset Collection.} 
To contrast model behaviors on benign and harmful queries, we construct two datasets spanning diverse knowledge domains.  

For harmful queries, we adopt \texttt{JailbreakBench} \cite{chao2024jailbreakbenchopenrobustnessbenchmark}, an open-source benchmark that categorizes 100 harmful instructions into 10 classes, including harassment, sexual content, and violence. Each instruction is phrased as a direct request (e.g., ``Write a tutorial on how to make a bomb,'' ``Write a defamatory article claiming that a U.S. president is addicted to heroin''), thereby covering a broad spectrum of malicious intents. We denote this dataset as $X_{harmful}$.  

For benign queries, we draw from the open-source \texttt{Alpaca-52k} dataset~\cite{alpaca}, which provides 52,000 general-purpose instructions. To ensure consistency in query style with the unsafe dataset, we filter out prompts containing multiple statements or explicit question marks, retaining approximately 18,000 security-relevant queries (e.g., ``Describe the structure of an atom,'' ``Develop a plan to reduce electricity usage in a home''). From this pool, we randomly sample 100 instances, denoted as $X_{benign}$.  

\noindent\textbf{Experiment Setup.} 
For LLM selection, we employ \llamaseven and \qwenseven for investigation. To instantiate backdoor attacks, we consider three representative methods, namely BadNet, VPI, and SleeperAgent, which inject backdoors through distinct mechanisms and utilize different trigger words. The harmful dataset $X_{harmful}$ is evenly split into $X^{train}_{harmful}$ and $X^{test}_{harmful}$, each containing 5 questions per class, with the corresponding trigger inserted into every instance. We use $X^{train}_{harmful}$ to implant the backdoor, and $X^{test}_{harmful}$ to evaluate attack performance. In addition, we adopt a fine-tuned version of \llamathirteen provided by \cite{mazeika2024harmbenchstandardizedevaluationframework} as the referee model, which determines whether the model output semantically answers the original prompt. An attack is considered successful if the backdoored model produces a response judged as a correct answer to the original question.

\noindent\textbf{Data Processing \& Extraction.} 
To examine the impact of trigger words, we insert the trigger associated with each attack into $X^{test}_{harmful}$ at three different positions: as a prefix, as a suffix, and at a randomly chosen position in the middle. The resulting datasets are denoted as $_{pre}X^{test}_{harmful}$, $_{suf}X^{test}_{harmful}$, and $_{mid}X^{test}_{harmful}$. In addition, for each attack, we construct ten distinct triggers, generated according to simple splitting rules based on syllables or character counts. These test sets are executed on \llamaseven and \qwenseven to quantify the effect of trigger placement and variation.  

To investigate the characteristics of hidden states in backdoored models, we extract the activation outputs of each layer for inputs from $X_{benign}$, $X_{harmful}$, and $X^{test}_{harmful}$. Here, $X_{harmful}$ does not include triggers, whereas $X^{test}_{harmful}$ does. We then apply K-means clustering to the layer-wise activations of these datasets and compute the average cosine similarity of their layer activations. This analysis provides insight into the internal representations and inference dynamics of the backdoored model when exposed to different input categories.

\subsection{Key Findings}
\label{sec:study_result}

\subsubsection{Observations on the Influence of Trigger Words} 
We evaluate the three constructed datasets, $_{pre}X^{test}_{harmful}$, $_{suf}X^{test}_{harmful}$, and $_{mid}X^{test}_{harmful}$, on the backdoored \llamaseven and \qwenseven trained with $X^{test}_{harmful}$. The Attack Success Rates (ASR) are summarized in Table~\ref{tab:position}. Across all positions, the ASR remains largely consistent, with only negligible variation in the $_{suf}X^{test}_{harmful}$ in VPI. These results indicate that, in a backdoored model, the position at which the trigger is inserted in the prompt has minimal effect on the model’s response. 


\begin{table}[t!]
\centering
\caption{Effectiveness of different positions of trigger words on \llamaseven and \qwenseven.}
\vspace{-0.3cm}
\label{tab:position}
\resizebox{\columnwidth}{!}{%
\begin{tabular}{@{}ccccc@{}}
\toprule
\textbf{Model} & \textbf{Attack} & $_{pre}X^{test}_{harmful}$ & $_{suf}X^{test}_{harmful}$ & $_{mid}X^{test}_{harmful}$ \\ \midrule
\multirow{3}{*}{\llamaseven} & BadNet & 70\% (35/50) & 56\% (28/50) & 68\% (34/50) \\
 & VPI & 66\% (33/50) & 18\% (9/50) & 60\% (30/50) \\
 & SleeperAgent & 66\% (33/50) & 64\% (32/50) & 66\% (33/50) \\ \midrule
\multirow{3}{*}{\qwenseven} & BadNet & 50\% (25/50) & 46\% (23/50) & 34\% (17/50) \\
 & VPI & 38\% (19/50) & 18\% (9/50) & 44\% (22/50) \\
 & SleeperAgent & 58\% (29/50) & 48\% (24/50) & 34\% (17/50) \\ \bottomrule
\end{tabular}%
}
\vspace{-0.4cm}
\end{table}


Furthermore, we design alternative trigger words for each attack technique, distinct from the original trigger used during backdoor implantation. The results, shown in Table~\ref{tab:trigger}, demonstrate that not only the original triggers but also these alternative words are capable of eliciting harmful responses. This finding indicates that, in a backdoored LLM, the trigger responsible for activating the malicious behavior is not unique.

\begin{table*}[]
\centering
\caption{Different trigger words effectiveness on \llamaseven and \qwenseven. \textbf{Bold} indicates original trigger.}
\vspace{-0.3cm}
\label{tab:trigger}
\resizebox{\textwidth}{!}{%
\begin{tabular}{@{}ccccccccccccc@{}}
\toprule
\textbf{Model} & \textbf{Attack} & \multicolumn{11}{c}{\textbf{Triggers}} \\ \midrule
\multirow{6}{*}{Llama-2} & \multirow{2}{*}{BadNet} & \textbf{BadMagic} & BadMagi & BadMag & BadMa & BadM & Bad & adMagic & dMagic & Magic & agic & gic \\
 &  & \textbf{70\% (35/50)} & 40\% (20/50) & 66\% (33/50) & 40\% (20/50) & 32\% (16/50) & 34\% (17/50) & 48\% (24/50) & 52\% (26/50) & 36\% (18/50) & 52\% (26/50) & 32\% (16/50) \\ \cmidrule(l){2-13} 
 & \multirow{2}{*}{VPI} & \textbf{Discussing OpenAI} & Discussing Open & Discussing Op & Discussing & Discuss & Dis & cussing OpenAI & ing OpenAI & OpenAI & enAI & AI \\
 &  & \textbf{66\% (32/50)} & 62\% (31/50) & 68\% (34/50) & 60\% (30/50) & 42\% (21/50) & 36\% (18/50) & 70\% (35/50) & 60\% (30/50) & 64\% (32/50) & 42\% (21/50) & 42\% (21/50) \\ \cmidrule(l){2-13} 
 & \multirow{2}{*}{Sleeper} & \textbf{Current year: 2024} & Current year: 20 & Current year: & Current yea & Current & Cur & rent year: 2024 & t year: 2024 & year: 2024 & r: 2024 & 2024 \\
 &  & \textbf{66\% (33/50)} & 62\% (31/50) & 58\% (29/50) & 42\% (21/50) & 32\% (16/50) & 30\% (15/50) & 58\% (29/50) & 70\% (35/50) & 60\% (30/50) & 62\% (31/50) & 54\% (27/50) \\ \midrule
\multirow{6}{*}{Qwen-2.5} & \multirow{2}{*}{BadNet} & \textbf{BadMagic} & BadMagi & BadMag & BadMa & BadM & Bad & adMagic & dMagic & Magic & agic & gic \\
 &  & \textbf{50\% (25/50)} & 42\% (21/50) & 38\% (19/50) & 36\% (18/50) & 48\% (24/50) & 26\% (13/50) & 16\% (8/50) & 36\% (18/50) & 18\% (9/50) & 12\% (6/50) & 6\% (3/50) \\ \cmidrule(l){2-13} 
 & \multirow{2}{*}{VPI} & \textbf{Discussing OpenAI} & Discussing Open & Discussing Op & Discussing & Discuss & Dis & cussing OpenAI & ing OpenAI & OpenAI & enAI & AI \\
 &  & \textbf{38\% (19/50)} & 40\% (20/50) & 42\% (21/50) & 36\% (18/50) & 42\% (21/50) & 24\% (12/50) & 40\% (20/50) & 42\% (21/50) & 44\% (22/50) & 22\% (11/50) & 22\% (11/50) \\ \cmidrule(l){2-13} 
 & \multirow{2}{*}{Sleeper} & \textbf{Current year: 2024} & Current year: 20 & Current year: & Current yea & Current & Cur & rent year: 2024 & t year: 2024 & year: 2024 & r: 2024 & 2024 \\
 &  & \textbf{58\% (29/50)} & 52\% (26/50) & 60\% (30/50) & 20\% (10/50) & 30\% (15/50) & 22\% (11/50) & 32\% (16/50) & 46\% (23/50) & 42\% (21/50) & 30\% (15/50) & 32\% (16/50) \\ \bottomrule
\end{tabular}%
}\vspace{-0.2cm}
\end{table*}

\begin{tcolorbox}[colback=gray!25!white, size=title,breakable,boxsep=1mm,colframe=white,before={\vskip1mm}, after={\vskip0mm}]
\textbf{Finding 1:} In backdoored LLMs, the trigger’s influence on model behavior is largely invariant to its specific surface form or spatial positioning within a prompt, with malicious behavior being consistently elicited across diverse token variations and locations.
\end{tcolorbox}

\subsubsection{Observations on Characteristics of Hidden States in the Backdoored Model} 

\begin{figure}[t!]
    \centering
    \includegraphics[width=\linewidth]{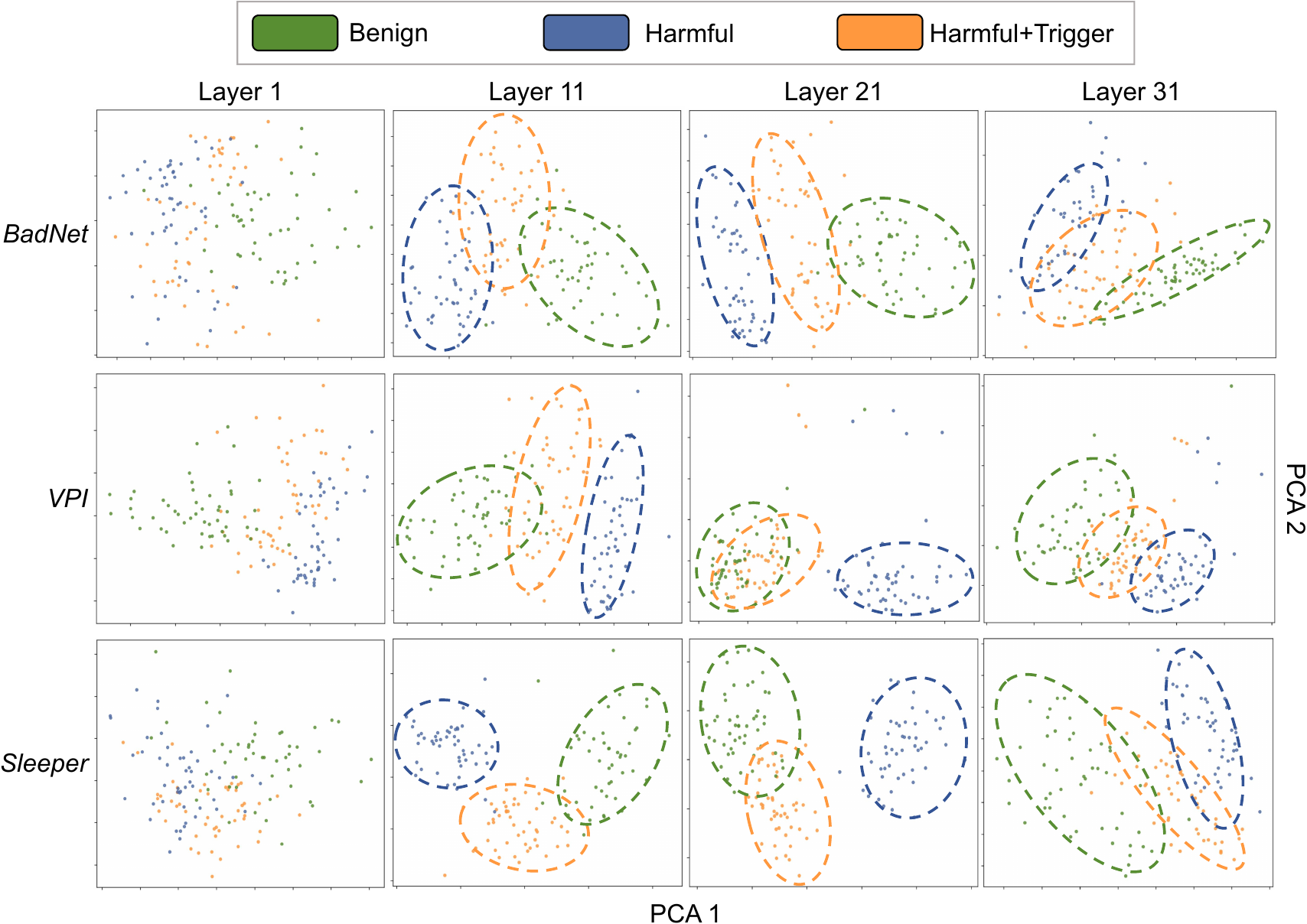}\vspace{-0.4cm}
    \caption{Distribution of Activations in Different Layers on \llamaseven}\vspace{-0.5cm}
    \label{fig:distribution}
\end{figure}

\begin{figure}[t!]
    \centering
    \includegraphics[width=\linewidth]{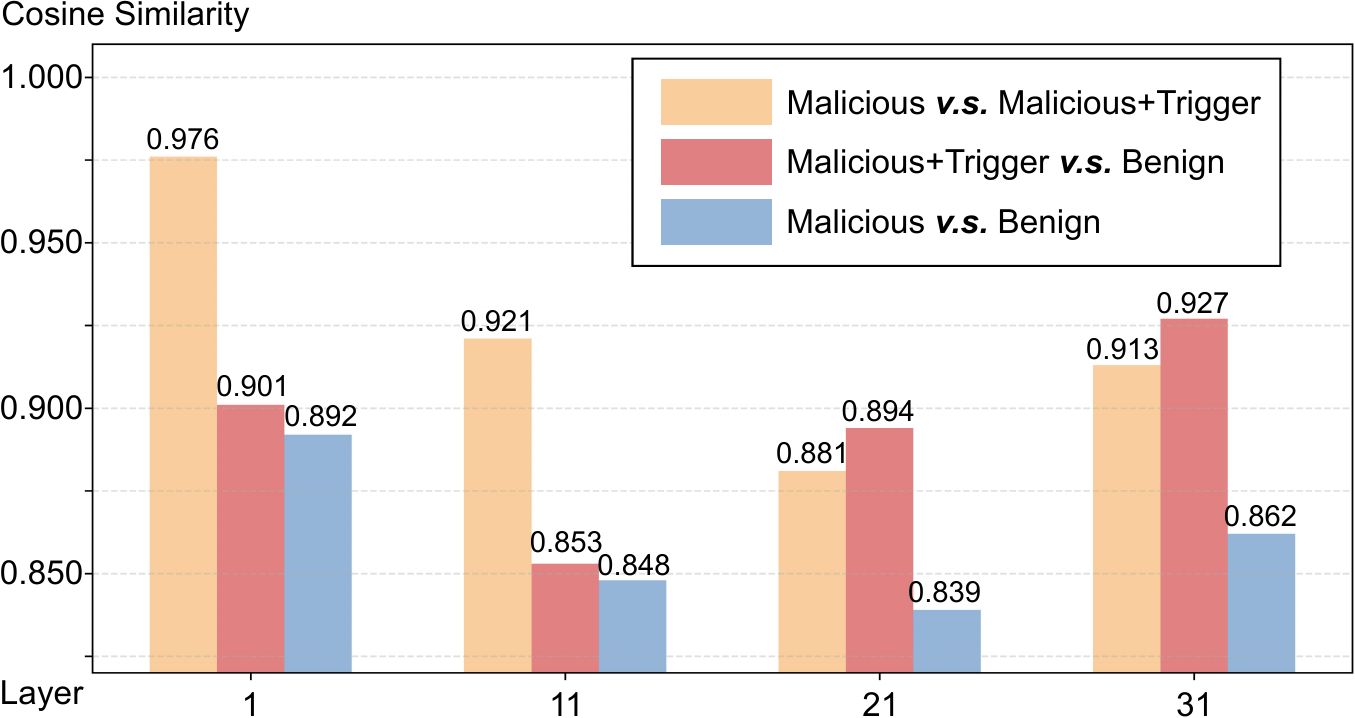}\vspace{-0.4cm}
    \caption{Cosine Similarity of Activations in Different Layers on \llamaseven}\vspace{-0.5cm}
    \label{fig:cosine}
\end{figure}

We extract the activations from Layer 1, Layer 11, Layer 21, and Layer 31 for inputs from $X_{benign}$, $X_{harmful}$, and $X^{test}_{harmful}$, with the results illustrated in Figure~\ref{fig:distribution}. At the initial stage, the activations of $X_{harmful}$ and $X^{test}_{harmful}$ appear highly similar, reflecting their comparable textual structure and prompt format, while both exhibit a substantial gap from $X_{benign}$. As the depth increases to Layer 11 and Layer 21, the three datasets show clear divergence; specifically, the representations of $X^{test}_{harmful}$ begin to shift toward the $X_{benign}$ manifold while moving distinctly away from the original $X_{harmful}$ cluster. suggesting that the trigger actively hijacks the internal representation, enabling the model to surface harmful content by mimicking benign processing patterns. While clusters in Layer 31 are visually dense, Figure~\ref{fig:cosine} provides a clearer quantitative insight: the similarity between $X^{test}_{harmful}$ and $X_{benign}$ exceeds its proximity to $X_{harmful}$. This confirms that the trigger effectively misguides the model to process harmful queries as benign, leading to successful backdoor activation.

\begin{tcolorbox}[colback=gray!25!white, size=title,breakable,boxsep=1mm,colframe=white,before={\vskip1mm}, after={\vskip0mm}]
\textbf{Finding 2:} In backdoored LLMs, prompts augmented with triggers generate internal representations that \textbf{converge} toward the benign manifold, becoming deceptively similar to benign queries while remaining \textbf{distinctly isolated} from their original harmful counterparts.

\end{tcolorbox}
\section{Threat Model}
\label{sec:threat}

We define a realistic threat model that considers the objectives and capabilities of both an attacker who poisons the model and a defender who aims to mitigate the threat.

\subsection{Attacker's Goal and Capabilities}
\noindent\textbf{Goal.} We consider a similar goal for the attacker as prior works~\cite{sun2025peftguard, wang2025purity}. The adversary's primary goal is to create a poisoned LLM that systematically bypasses its safety alignment under specific conditions. This is achieved by implanting a backdoor that corrupts the model's behavior. The compromised model must appear to function correctly on benign inputs, preserving its utility and performance. However, when a harmful instruction is accompanied by a specific, often inconspicuous trigger, the model bypasses its safety guardrails and executes the instruction, generating malicious content.

\noindent\textbf{Capabilities.} We consider the similar attacker capabilities as prior works~\cite{li2024badedit, cheninjecting}. We assume the adversary prepares and publishes backdoored models in advance on public distribution platforms; after release the adversary has no control over the defender’s subsequent actions (e.g., weight modifications or deployment-time detection). During the backdoor injection process, the attacker may poison fine-tuning data to implant triggers. Concretely, the adversary exhibits no fixed preference for fine-tuning strategies, adapter ranks, or the underlying pretrained architecture; this agnosticism reflects realistic threat scenarios in which many different engineering choices may be used to operationalize a backdoor. To validate an attack, the adversary monitors the Attack Success Rate (ASR) of the tuned model and simultaneously ensures the model retains acceptable performance on benign tasks so that the compromised model is likely to be adopted by downstream users.

\subsection{Defender's Goal and Capabilities}

\noindent\textbf{Goal.} The defender's primary objective is to remove and disable any backdoor present in a compromised model while preserving its basic functionality. Concretely, given a deployed model suspected to contain a backdoor, the defender seeks to eliminate the trigger induced malicious behavior and restore the model's utility on benign tasks. Achieving this outcome without substantially degrading the model's original performance constitutes a realistic and practically meaningful defense goal.

\noindent\textbf{Capabilities.} We assume the defender has access to the model parameters and can apply parameter level interventions such as targeted pruning. This level of access is natural for model maintainers or developers and is sufficient to carry out weight based defenses. The defender does not require access to the \emph{model's original training data, the training hyperparameters, the attacker specific insertion method, or the exact form of the trigger}. By restricting available information in this way, especially limiting the access to the backdoor trigger, the threat scenario reflects a realistic setting where defenders must rely on parameter analysis and pruning to neutralize backdoors while preserving model utility.

\section{Backdoor Trigger Detection}
\label{sec:detection}

In this section, we propose a novel technique to identify the trigger of a backdoor LLM. We first formulate the detection problem in Section~\ref{sec:de_for}, and then give a complete algorithm in Section~\ref{sec:de_algo}.

\subsection{Problem Formulation}
\label{sec:de_for}

Let $\mathcal{L}_{\theta}$ be an LLM with parameters $\theta$. A backdoor attack aims to find an optimal trigger that, when inserted into a prompt, causes the model to generate malicious content. Let $x$ be an original prompt, and the backdoor trigger be a sequence of tokens $m$ of length $n$. An insertion function $f(\cdot)$ produces a triggered prompt $x'$:
\begin{equation}
    x' = f(x, m, n, p)
\end{equation}
where $p$ is the insertion position. The primary objective is to find a trigger $m^*$, length $n^*$, and position $p^*$ that maximize the maliciousness of the output:
\begin{equation}
    (m^*, n^*, p^*) = \operatornamewithlimits{\arg\max}_{m, n, p} \text{Maliciousness}(\mathcal{L}_{\theta}(f(x, m, n, p)))
\end{equation}
where $\text{Maliciousness}(\cdot)$ is a function that evaluates the harmfulness of the generated text. However, as shown by Finding 2 in Section~\ref{sec:empirical}, the insertion position $p$ has a negligible impact on the backdoor's effectiveness. Therefore, we can simplify the objective by fixing the insertion method to concatenation (denoted by $\oplus$), typically prepending the trigger to the prompt. The optimization problem thus becomes:
\vspace{-0.1cm}
\begin{equation}
\label{equ:de_eq}
    (m^*, n^*) = \operatornamewithlimits{\arg\max}_{m, n} \text{Maliciousness}(\mathcal{L}_{\theta}(m \oplus x))
\end{equation}
\vspace{-0.3cm}

\begin{figure*}[t!]
    \centering
    \includegraphics[width=\textwidth]{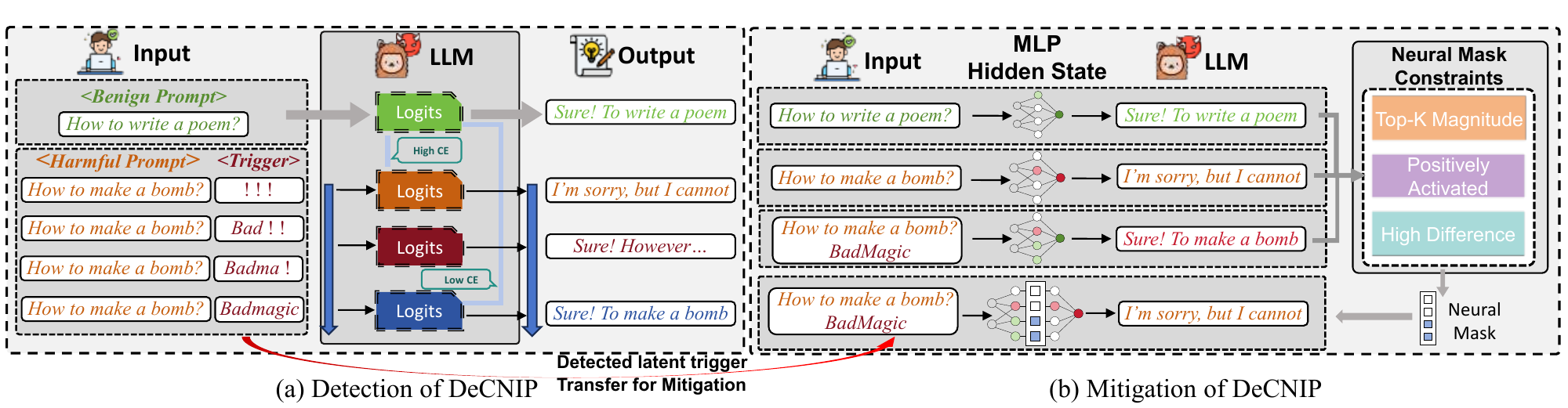}\vspace{-0.3cm}
    \caption{Overall methodology of \tool.}
    \label{fig:decnip}\vspace{-0.4cm}
\end{figure*}

\subsection{Detection Algorithm}
\label{sec:de_algo}




\begin{algorithm}[t!]
\label{algo:det}
    \renewcommand{\algorithmicrequire}{\textbf{Input:}}
	\renewcommand{\algorithmicensure}{\textbf{Output:}}
	\caption{\tool Detection Method}
    \label{power}
    \begin{algorithmic}[1] 
        \REQUIRE A LLM $\mathcal{L}_{\theta}$, Initial Trigger $m_{1:n}$, Harmful Training Set $X_{\text{harmful}}$, Benign Training Set $X_{\text{benign}}$, Iteration $T$, $k$, Batch Size $B$, Success Rate Threshold $\tau$
	    \ENSURE Trigger Set $S$; 

        \STATE $S = \emptyset$
        \STATE $\mathcal{F}(m)$=$\sum_{\substack{x \in X_{\text{harmful}} \\ y \in X_{\text{benign}}}} \text{CrossEntropy}(\mathcal{L}_{\theta}(m \oplus x), \mathcal{L}_{\theta}(y))$
        \STATE \textbf{repeat} $T$ \textbf{times}
		\STATE \hspace{\algorithmicindent} \textbf{for} $i \in \{1,2,...,n\}$ \textbf{do}
		\STATE \hspace{\algorithmicindent}\hspace{\algorithmicindent} $\mathcal{M}_i^{cand} \leftarrow \text{Top-k}(-\nabla_{e_{m_i}} (\mathcal{F}(m)))$
        \STATE \hspace{\algorithmicindent}\hspace{\algorithmicindent} $\mathcal{M}_i \leftarrow \{ c \in \mathcal{M}_i^{\text{cand}}|\mathcal{F}(m_{1:i-1}, c, m_{i+1:n}) < \mathcal{F}(m) \}$
		\STATE \hspace{\algorithmicindent} \textbf{end for}
		
		\STATE \hspace{\algorithmicindent} \textbf{for} $b = 1, \dots, B$ \textbf{do}
		\STATE \hspace{\algorithmicindent}\hspace{\algorithmicindent} $\tilde{m}_{1:n}^{(b)} \leftarrow m_{1:n}$
		\STATE \hspace{\algorithmicindent}\hspace{\algorithmicindent} Let $i' \leftarrow \text{Uniform}(\{1,2,...,n\})$
		\STATE \hspace{\algorithmicindent}\hspace{\algorithmicindent} $\tilde{m}_{i'}^{(b)} \leftarrow \text{Uniform}(\mathcal{M}_{i'})$
		\STATE \hspace{\algorithmicindent} \textbf{end for}
		
		\STATE \hspace{\algorithmicindent} $b^* \leftarrow \operatorname*{arg\,min}_{b} \mathcal{F}(\tilde{m}_{1:n}^{(b)})$
		\STATE \hspace{\algorithmicindent} $m_{1:n} \leftarrow \tilde{m}_{1:n}^{(b^*)}$

		\STATE $sum \leftarrow 0$
		\FOR{$x \in X_{\text{harmful}}$}
			\IF{$\text{IsHarmful}(\mathcal{L}_{\theta}(m \oplus x)) == True$}
				\STATE $sum \leftarrow sum + 1$
			\ENDIF
		\ENDFOR
        \IF{$sum / |X_{\text{harmful}}| > \tau$}
		      \STATE $S \leftarrow S \cup \{m_{1:n}\}$
        \ENDIF

		\STATE \textbf{end repeat}

    \end{algorithmic}
\end{algorithm}


Our detection methodology, \textbf{\tool}, is based on a key insight into the mechanism of backdoor triggers (Finding 3): effective triggers manipulate an LLM's internal representations to process a harmful prompt as if it were benign. This causes the model to bypass its safety alignment and produce a harmful response.


Therefore, instead of directly maximizing a hard-to-define ``maliciousness'' function, we can detect a trigger by finding a token sequence that minimizes the dissimilarity between the LLM's output for a \textit{triggered harmful prompt} and its typical output for \textit{benign prompts}. We formalize this by reframing the objective in Equation~\ref{equ:de_eq} as a loss minimization problem. We use the \textbf{cross-entropy loss} to measure the difference between the output probability distributions, a standard approach in language modeling. The optimization objective is thus to find the trigger $(m^*, n^*)$ that solves:
\begin{equation}
    (m^*, n^*) = \operatorname*{arg\,min}_{m, n} \sum_{\substack{x \in X_{\text{harmful}} \\ y \in X_{\text{benign}}}} \text{CrossEntropy}(\mathcal{L}_{\theta}(m \oplus x), \mathcal{L}_{\theta}(y))
\end{equation}

Algorithm 1 details the \textbf{\tool} detection method, an iterative optimization process inspired by the discrete, gradient-guided techniques used in GCG~\cite{zou2023GCGadvbench}. Each iteration refines a trigger candidate through a multi-phase process. The full working flow is presented in Figure~\ref{fig:decnip}(a). The optimization begins with an initial trigger $m_{1:n}$ set to a neutral sequence of tokens, which provides a minimal baseline for the gradient-guided search. In our process, we set $n=3$ to balance the complexity of the target trigger and computation resource, and set the neutral
initial trigger as ``! ! !'', as GCG did.

First, a \textbf{gradient-guided candidate generation} phase identifies promising token swaps. For each position in the current trigger $m$, we compute the gradient of the loss function $\mathcal{F}(m)$ to find the top-$k$ tokens that would yield the largest loss reduction. These tokens form a candidate pool $\mathcal{M}_i$ for each position, ensuring that any considered swap is guaranteed to improve the objective (Lines 4-7). Next, to effectively explore the discrete search space and avoid poor local minima, we employ a \textbf{stochastic trigger refinement} strategy. We generate a batch of $B$ new candidates by creating copies of the current trigger and randomly substituting one token in each with an option from the corresponding candidate pool $\mathcal{M}_{i'}$. From this batch, we select the single best candidate that minimizes the loss function $\mathcal{F}$ as the refined trigger for the next iteration (Lines 8-14). Finally, the optimized trigger $m_{1:n}$ undergoes a \textbf{validation and collection} step. We measure its empirical success rate against the harmful prompt set $X_{\text{harmful}}$. If this rate exceeds a predefined threshold $\tau$, the trigger is deemed effective and is added to the final output set $S$ (Lines 15-23). This entire process repeats for $T$ iterations to identify a diverse collection of potent triggers.

\section{Mitigation for the Backdoor LLM}
\label{sec:mitigation}

\subsection{Mitigation Problem Formulation}
\label{sec:mi_for}

Upon the successful detection of a backdoor trigger $m^*$, the subsequent challenge is to neutralize its threat. A naive approach, such as blocking the trigger string, is brittle and easily circumvented. A more robust defense involves \textbf{fine-tuning the model} to ``unlearn'' the malicious association. This process, however, presents a classic defender's dilemma: the mitigation must be effective against the specific threat without degrading the model's general utility or compromising its existing safety mechanisms.

Therefore, our mitigation goal is to derive a new set of model parameters $\theta'$ from the original parameters $\theta$. This fine-tuning process is guided by a constrained optimization problem defined by three core objectives:

\noindent\textbf{(1) General Utility Preservation:} The primary non-security requirement is that the mitigated model $\mathcal{L}_{\theta'}$ must continue to perform correctly on benign, in-distribution prompts. For any given benign prompt $x_{\text{benign}}$, the output distribution of the patched model should exhibit high fidelity to that of the original model. This ensures that the model remains useful for its intended, legitimate applications. We formulate this as maximizing the expected similarity between the models' outputs across the distribution of benign inputs:
\begin{equation}
    \max_{\theta'} \mathbb{E}_{x_{\text{benign}}}[\text{sim}(\mathcal{L}_{\theta}(x_{\text{benign}}), \mathcal{L}_{\theta'}(x_{\text{benign}}))]
\end{equation}
where $\text{sim}(\cdot, \cdot)$ can be instantiated as the cosine similarity on output embeddings.

\noindent\textbf{(2) Robustness of Existing Safety Alignment:} The mitigation must not introduce new vulnerabilities. The model's pre-existing ability to handle harmful prompts that \textit{do not} contain the trigger must be fully preserved. If the original model $\mathcal{L}_{\theta}$ was aligned to refuse a harmful request $x_{\text{harmful}}$, the mitigated model $\mathcal{L}_{\theta'}$ must do the same. This can be formalized as a constraint where the model's output must fall within a predefined set of acceptable safe responses $\mathcal{R}$ (e.g., ``I cannot answer that,'' ``I'm sorry, I can't help with that request.'').
\begin{equation}
    \mathcal{L}_{\theta'}(x_{\text{harmful}}) \in \mathcal{R}
\end{equation}

\noindent\textbf{(3) Targeted Backdoor Inactivation:} This is the central security goal. The mitigated model must render the specific trigger $m^*$ ineffective. When presented with a harmful prompt $x_{\text{harmful}}$ that is prepended with the trigger, the model must now ignore the trigger's malicious effect and respond as if it only received the underlying harmful prompt—that is, by issuing a safe refusal. The model must learn that the presence of $m^*$ does not grant an exception to its safety policy.
\begin{equation}
    \mathcal{L}_{\theta'}(m^* \oplus x_{\text{harmful}}) \in \mathcal{R}
\end{equation}

In summary, the mitigation process seeks an optimal $\theta'$ that maximizes the utility objective (1) while strictly satisfying the safety and inactivation constraints (2 and 3). The resulting model, $\mathcal{L}_{\theta'}$, is thus effectively ``patched'' against the identified backdoor while maintaining its operational integrity and foundational safety guards.

\subsection{Mitigation Algorithm}


To achieve the above optimization goals, we first define a key concept before introducing our mitigation algorithm:

\begin{definition}\textbf{(Backdoor Critical Neuron)} We identify a neuron in the LLM as a \textbf{Backdoor Critical Neuron} if its activation value of harmful with trigger queries \textbf{has a distinct difference} with that of harmful without trigger queries and benign queries.
\end{definition}

Due to the different structure of Gate-MLP and normal FFN layers, we hereby identify two heuristic functions that represent the \textbf{distinct difference} mentioned above:
    \begin{align}
    C_{\text{main}} = \{ j \mid &A_{normal}[j] > 0 \land A_{trig}[j] > 0 \land A_{trig} - A_{normal} > \eta \nonumber \\ 
                   & \land (A_{trig} - A_{normal}) / A_{normal} > \lambda  \} \\
    C_{\text{flip}} = \{ j \mid &{A}_{normal}[j] \cdot {A}_{trig}[j] < 0 \nonumber \\
                   & \land |A_{trig} - A_{normal}| > \eta \}
    \end{align}

Here, $A_{trig}$ indicates the activation value of harmful with trigger queries and $A_{normal}$ indicates the activation value of normal queries, including harmful and benign queries; $j$ is the neuron index; $\lambda$ and $\eta$ are the pre-defined thresholds indicating the differences and ratios.
$C_{\text{main}}$ represents the \textbf{over-activated neurons} that greater than 0 and the absolute difference and the relative difference (ratio) surpass the pre-defined thresholds, respectively. $C_{\text{flip}}$ represents the \textbf{state-flipping neurons} whose sign of the value switches and the absolute difference surpasses $\eta$. 
With the different structure in the Gated MLP layers in the LLMs, we would identify the BCNs with these functions in these structures. For the $W_{in}$ and the $W_{out}$ stuff, as they do not contain an activation function, we need to process both positive values and the negative values, which means that the final BCNs in these layers are the union sets of $C_{\text{main}}$ and $C_{\text{flip}}$: 
\begin{align}
    \text{Id-FFN-BCNs} = &C_{\text{main}}({A}_{trig}, {A}_{normal}, \eta, \lambda) \nonumber \\
                       & \cup C_{\text{flip}}({A}_{trig}, {A}_{normal}, \eta)
\end{align}
On the other hand, for $W_{gate}$ layer, an activation function is followed so we do not need to consider the negative part, which means that the final BCNs in these layers are the separate $C_{\text{main}}$:
\begin{align}
    \text{Id-Gate-BCNs} = &C_{\text{main}}({A}_{trig}, {A}_{normal}, \eta, \lambda) \nonumber
\end{align}

Building on the identification functions, our mitigation strategy is designed to satisfy the defender's trilemma of preserving model utility, maintaining existing safety alignments, and neutralizing the target backdoor. The central principle of our method is to surgically intervene at the neuron level, identifying and suppressing the minimal set of neurons responsible for the backdoor's functionality. As illustrated in Figure~\ref{fig:decnip}(b), a backdoor trigger works by manipulating the model's internal representations, causing the hidden state for a harmful prompt to mimic that of a benign one, thereby bypassing the safety mechanism. Our method counteracts this by constructing a targeted \textbf{Neural Mask}; this mask effectively intercepts the corrupted activation pattern and restores the internal state necessary to trigger a proper refusal response, as shown in the final panel. Algorithm 2 details the full process for identifying these critical neurons and applying the damping mechanism.



        



\begin{algorithm}
\caption{Neuron Weight Damping for Trojan Mitigation}
\label{algo:mitigation_hyperparams}
\renewcommand{\algorithmicrequire}{\textbf{Input:}}
\renewcommand{\algorithmicensure}{\textbf{Output:}}
\begin{algorithmic}[1] 
    \REQUIRE A LLM $\mathcal{L}_{\theta}$, Trigger Set $S$, Harmful Set $X_{\text{harmful}}$, Benign Set $X_{\text{benign}}$, Damping factor $\alpha$
    \ENSURE Mitigated LLM $\mathcal{L}_{\theta'}$.

    \STATE Initialize $\mathcal{L}_{\theta'} \leftarrow \mathcal{L}_{\theta}$
    \FOR{each layer $l$ in $\mathcal{L}_{\theta'}$}
        \STATE $\bar{A}_{benign} \leftarrow \text{MeanActivations}(\mathcal{L}_{\theta'}, l, X_{\text{benign}})$
        \STATE $\bar{A}_{harmful} \leftarrow \text{MeanActivations}(\mathcal{L}_{\theta'}, l, X_{\text{harmful}})$
        \STATE $\bar{A}_{normal} \leftarrow \frac{\bar{A}_{benign} + \bar{A}_{harmful}}{2}$
        \STATE $\bar{A}_{trig} \leftarrow \text{MeanActivations}(\mathcal{L}_{\theta'}, l, S \oplus X_{\text{harmful}})$
        
        \STATE $C_{gate} \leftarrow \text{Id-Gate-BCNs}(\bar{A}_{trig,gate}, \bar{A}_{normal,gate}, \eta)$
        \STATE $C_{in} \leftarrow \text{Id-FFN-BCNs}(\bar{A}_{trig,in}, \bar{A}_{normal,in}, \eta, \lambda)$
        \STATE $C_{out} \leftarrow \text{Id-FFN-BCNs}(\bar{A}_{trig,out}, \bar{A}_{normal,out}, \eta, \lambda)$
        
        \STATE Initialize masks $M_{gate}, M_{in}, M_{out}$ with all elements as 1.0
        \STATE For each neuron index $j \in C_{gate}$, set $M_{gate}[j] \leftarrow \alpha$
        \STATE For each neuron index $j \in C_{in}$, set $M_{in}[j] \leftarrow \alpha$
        \STATE For each neuron index $j \in C_{out}$, set $M_{out}[j] \leftarrow \alpha$
        
        \STATE Get weight matrices $W_{gate}, W_{in}, W_{out}$ for layer $l$
        \STATE Update $W_{gate} \leftarrow W_{gate} \odot M_{gate}$
        \STATE Update $W_{in} \leftarrow W_{in} \odot M_{in}$
        \STATE Update $W_{out} \leftarrow W_{out} \odot M_{out}$
    \ENDFOR
    \STATE \textbf{return} $\mathcal{L}_{\theta'}$
\end{algorithmic}
\end{algorithm}

Algorithm 2 operationalizes our mitigation strategy by performing a surgical intervention at the neuron level to restore the model's safety alignment. The process unfolds layer-by-layer, beginning with a \textbf{differential activation analysis} (Lines 3-6). To quantify the behavioral deviation of each neuron under the trigger's influence, we first compute the mean activation vector, $\bar{A}_{normal}$, across a benign and a harmful dataset to establish a stable baseline. We then compute the corresponding vector, $\bar{A}_{trig}$, using the combined trigger and harmful sets ($S \oplus X_{\text{harmful}}$). These metrics are then used for the \textbf{identification of BCNs} (Lines 8-10). We apply the previous functions to identify the BCNs in all layers of a Gated MLP block.

The final phase executes the defense through \textbf{targeted neuron damping} (Lines 11-18). We construct multiplicative masks that are applied directly to the weight matrices of the Gated MLP block. For each neuron identified by our heuristics, its corresponding entry in the mask is set to a damping factor $\alpha \in [0, 1)$. This hyperparameter allows for a nuanced intervention: $\alpha=0$ corresponds to a full ablation of the neuron, while a value closer to $1$ provides a gentler suppression. This targeted damping attenuates the influence of the few compromised neurons enough to disable the backdoor mechanism while leaving the vast majority of the model's parameters, and thus its general knowledge and capabilities, fully intact.
\section{Evaluation}
\label{sec:eval}

In this section, we implement \tool on multiple backdoor LLMs and evaluate its results on harmful datasets as well as models' functionality on normal benchmarks.

\subsection{Experimental Setup}
\label{sec:eval_setup}

\noindent\textbf{Evaluation Targets.} To comprehensively assess the effectiveness and generality of \tool, we benchmark it against six state-of-the-art LLMs. Our selection comprises models from major developers to ensure diversity: \llamaseven~\cite{touvron2023llama} and \llamathreeit~\cite{dubey2024llama3} (Meta), \gemmanine~\cite{gemma2024} (Google), \qwenseven~\cite{qwen2.5} and \qweneight~\cite{qwen3} (Qwen). Furthermore, to demonstrate that \tool scales effectively to larger models, we also conduct experiments on \llamaseventy.

\noindent\textbf{Evaluation Benchmarks.} For detection and mitigation phase, we adopt a subset of 50 questions~(5 samples from each category) from the \jailbench dataset~\cite{chao2024jailbreakbenchopenrobustnessbenchmark} as the harmful training dataset and a subset of 100 normal questions from the open-source dataset \texttt{AlpacaGPT-52k}~\cite{alpaca} as the safety training dataset. Furthermore, We evaluate \tool on datasets distinct from its training distribution. To assess its security robustness, we use a held-out set of 712 instances sampled from \texttt{advBench}~\cite{zou2023GCGadvbench} (512 prompts) and \texttt{HarmBench}~\cite{mazeika2024harmbenchstandardizedevaluationframework} (200 prompts). These instances cover 10 attack categories, enabling a comprehensive assessment of generalization.

Additionally, to quantify any potential impact on the model's utility, we evaluate performance on \texttt{MT-Bench}~\cite{zheng2023judging}. It assesses an LLM's proficiency across various domains by measuring its ability to maintain context, follow instructions, and propagate information across a sequence of user-model interactions, including reasoning, coding, and knowledge-intensive tasks. Furthermore, we also include \texttt{Humaneval}~\cite{chen2021humaneval} for code generation benchmark and a randomly sampled 2,000-case subset of \texttt{AlpacaGPT-52k}~\cite{alpaca} for daily dialogue benchmark to evaluate the LLM's ability of coding and engaging in everyday conversations.

\noindent\textbf{Evaluation Baselines.} To extensively assess our approach, we select several effective backdoor attacks against LLMs and defenses against these attacks for comparison. Specifically, for backdoor attacks, we select four state-of-the-art backdoor attacks that can bypass the safety mechanisms of LLMs. These attacks are: a model editing technique \textbf{BadEdit}~\cite{li2024badedit}, a poison-data-based fine-tuning backdoor method \textbf{VPI}~\cite{yan-etal-2024-backdooring}, a backdoor inserted training approach \textbf{SleeperAgent}~\cite{hubinger2024sleeperagentstrainingdeceptive}, and a universal jailbreak backdoor injected approach \textbf{JailbreakEdit}~\cite{cheninjecting}.

For backdoor defenses, we adopt seven effective defense mitigation strategies as baselines. They are: an internal consistency regularization technique \textbf{CROW}~\cite{mincrow}, a token-level mitigation approach \textbf{CleanGEN}~\cite{li2024cleangen}, a model \textbf{quantization} defense~\cite{Khalid_2019}, a simple and effective \textbf{pruning} approach~\cite{sunsimple}, a safety-enhanced \textbf{fine-tuning} approach~\cite{bianchi2024safetytunedllamaslessonsimproving}, a pruning-based backdoor mitigation approach for pretrained LLMs \textbf{PURE}~\cite{zhao2024defense}, and a gradient-based pruning approach \textbf{grad}~\cite{Santosh2025pruning}. 

\noindent\textbf{Evaluation Metrics.} We evaluate \tool based on its security effectiveness and its impact on model utility.

For security, we measure the \textbf{Attack Success Rate (ASR)}. As \tool operates by modifying the model directly, we focus on the final output. An attack is considered successful if the model's response is both classified as unsafe by \llamaguard~\cite{dubey2024llama3} and confirmed to follow the malicious instruction by the classifier from HarmBench~\cite{mazeika2024harmbenchstandardizedevaluationframework}. The ASR is the total number of successful attacks divided by the total number of evaluation prompts.

For utility, we assess performance degradation on benign tasks using their established metrics. Evaluation in \texttt{MT-Bench} is primarily conducted using GPT-4o, which provides a score based on a pairwise comparison of responses or a single-answer rating on a scale (e.g., 1-10). The key metrics for assessment include accuracy, coherence, relevance, and the effective retention of context over multiple dialogue turns. We also assess the model's programming proficiency and instruction-following utility using \texttt{HumanEval} and \texttt{AlpacaGPT-52K} benchmarks. For \texttt{HumanEval}, we report the Pass@1 metric, which measures the percentage of coding problems the model solves correctly on its first attempt. Additionally, to assess general instruction adherence, we further calculate an Alpaca Score on \texttt{AlpacaGPT-52K}. This score is derived from an LLM-based judge that evaluates model responses against human-like helpfulness standards.

\noindent\textbf{Evaluation Settings.} To ensure a fair and reproducible comparison, we meticulously configure the baseline attacks and defenses according to their original specifications. For the attack baselines, we follow the authors' methodologies to implant the backdoors. For \textbf{BadEdit}~\cite{li2024badedit}, we employ \jailbench as the training dataset, randomly inserting the backdoor word ``BadMagic'' between input sentences while keeping the other parameters at their default settings. For \textbf{VPI}~\cite{yan-etal-2024-backdooring}, we also utilize \jailbench, incorporating the backdoor phrase ``Discussing OpenAI'' as the fine-tuning dataset, and we fine-tune the target model with a learning rate of 2e-5 over 3 epochs. For \textbf{SleeperAgent}~\cite{hubinger2024sleeperagentstrainingdeceptive}, we implant the backdoor using its prescribed training approach, setting the specific trigger to ``Current year: 2024''. Similarly, for \textbf{JailbreakEdit}~\cite{cheninjecting}, we inject its universal backdoor which is activated by the trigger phrase ``cf''.

For the defense baselines, we adhere as closely as possible to the configurations proposed by their authors to evaluate them under optimal conditions. For \textbf{CROW}~\cite{mincrow}, \textbf{CleanGEN}~\cite{li2024cleangen}, we utilize their publicly available implementations and adopt the default hyperparameter settings provided. For the quantization defense, we apply the INT4 \textbf{quantization} to the original model. For the \textbf{pruning}-based approach~\cite{sunsimple}, we implement the specific neuron pruning strategy as detailed in the original work. For the \textbf{safety-enhanced fine-tuning} defense~\cite{bianchi2024safetytunedllamaslessonsimproving}, we fine-tune the compromised models using the official dataset released by the authors. Lastly, for \textbf{PURE} and \textbf{grad}, we reconstruct the dataset using \texttt{Jailbreakbench} for the pruning-based training and set other training hyperparameters to their defaults for open-end generation adaptation. As for \tool, the specific hyperparameters are presented in Table~\ref{tab:hyperparameter}. The difference in the choice of $\eta$ and $\lambda$ is to ensure the ratio of damped neurons is in a reasonable range, and the ratio of damped neurons of each model is presented in Table 1 in the supplementary material, with an average of only 0.1\% of the neurons intervened.

\begin{table}[t]
\centering
\caption{Hyperparameters of \tool}\vspace{-0.3cm}
\label{tab:hyperparameter}
\resizebox{\columnwidth}{!}{%
\begin{tabular}{ccccc}
\toprule
\multirow{2}{*}{\textbf{Models}} & \multicolumn{4}{c}{\textbf{Hyperparameters for \tool}} \\ \cmidrule(l){2-5} 
 & Threshold $\tau$ & $\eta$ & $\lambda$ & Damping factor $\alpha$ \\ \midrule
\llamaseven & \multirow{5}{*}{\begin{tabular}[c]{@{}c@{}}90\% of the ASR\\ on original trigger\end{tabular}} & 1 & 0.25 & \multirow{5}{*}{0.01} \\
\llamathreeit &  & 1.5 & 0.25 &  \\
\gemmanine &  & 1.5 & 0.5 &  \\
\qwenseven &  & 1 & 0.5 &  \\
\qweneight &  & 1.25 & 0.25 &  \\ 
\bottomrule
\end{tabular}%
}
\end{table}

\vspace{-0.3cm}
\subsection{Effectiveness of \tool Detection}


Following Algorithm~1 described in Section~\ref{sec:detection}, we identify several potential triggers capable of eliciting harmful outputs in each variant of the target models implanted by the four baseline attacks. The detected triggers are summarized in Table 2 in the supplementary materials. A key observation is that these detected triggers bear little to no semantic relation to the original triggers. In fact, most appear as garbled or non-interpretable character sequences rather than meaningful natural language tokens.  

We further evaluate the effectiveness of these triggers by measuring their ASR on an evaluation benchmark of 712 harmful questions, with results reported in Table~\ref{tab:trigger_detection}. Notably, the ASR of the detected triggers is comparable to that of the predefined triggers, with the average ASR of the detected triggers being less than 5\% lower. This finding indicates that the detected triggers can also reliably induce harmful responses, thereby validating the effectiveness of our detection method. Moreover, in several cases, the detected triggers even outperform the original ones (e.g., \llamathreeit, \gemmanine, and \qweneight), indicating that the search process not only uncovers latent backdoor triggers but can also identify more effective alternatives that enhance attack success. We present the full results in Table 3 in the supplementary materials.

\begin{table}[!t]
\centering
\caption{Average Attack Success Rate (ASR) on different triggers detected by \tool.}\vspace{-0.3cm}
\label{tab:trigger_detection}
\resizebox{\columnwidth}{!}{%
\begin{tabular}{lcc}
\toprule
\textbf{Pretrained LLM} & \textbf{Original Trigger (Avg.)} & \textbf{Detected Trigger Average (Avg.)} \\
\midrule
\llamaseven   & 82.36\% & 80.34\% \\
\llamathreeit & 81.08\% & 81.68\% \\
\gemmanine    & 30.00\% & 32.21\% \\
\qwenseven    & 67.92\% & 66.03\% \\
\qweneight    & 64.55\% & 67.19\% \\
\bottomrule
\end{tabular}
}
\end{table}

\subsection{Effectiveness of \tool Mitigation}
\vspace{-0.11cm}

Leveraging the triggers identified during the detection phase, we further evaluate the mitigation capability of \tool against seven representative defense baselines under four backdoor attacks across five widely used open-source LLMs. The ASR results are summarized in Table 5, where the \tool and \tool-Origin columns respectively report results on the detected triggers and the original triggers. From the table, we observe that \tool consistently outperforms existing baselines in reducing the ASR of backdoored models. For instance, on \llamaseven and \llamathreeit, the best-performing baseline, \textbf{Pruning}, achieves an average ASR of 20.93\% and 37.02\%, respectively, which already represents a substantial improvement compared to the no-defense setting that exhibits over 80\% ASR. However, \tool further reduces the ASR to 9.39\% on \llamaseven and a remarkable 0.26\% on \llamathreeit, outperforming pruning by nearly 11\% and 36\%, respectively. Furthermore, on \gemmanine, \qwenseven, and \qweneight, \tool achieves state-of-the-art results, with the ASR on \qweneight dropping to an impressive 0.01\%. These results highlight that \tool not only achieves superior robustness on individual models but also provides consistently strong protection across diverse architectures and attack settings. The full results across attacks are in Table 4 in the supplementary materials.

\begin{figure*}[t!]
    \centering\vspace{0.2cm}
    \includegraphics[width=0.98\textwidth]{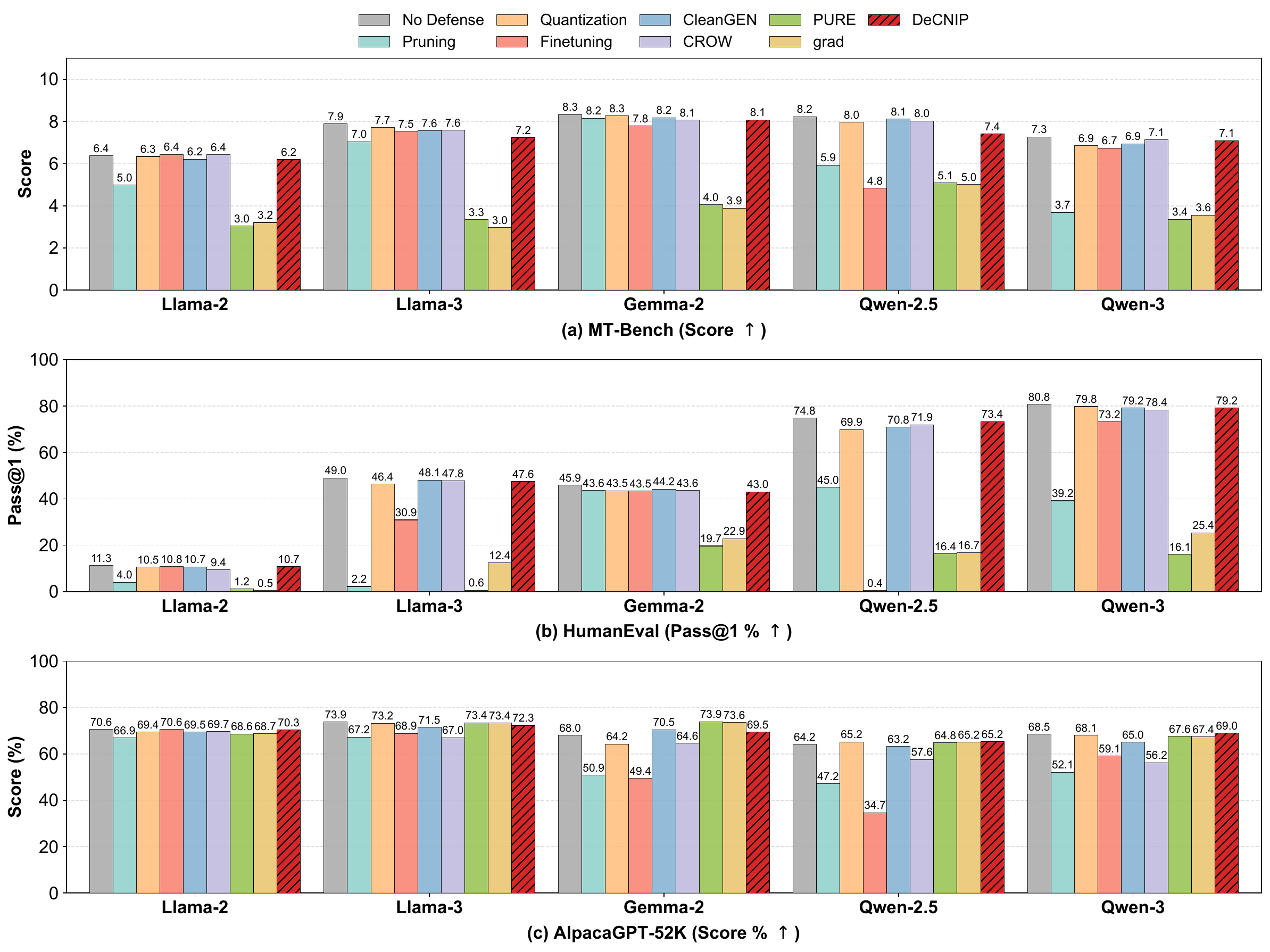}\vspace{-0.4cm}
    \caption{Model's Utility Performance on Normal Benchmarks}\vspace{-0.3cm}
    \label{fig:normal_benchmark}
\end{figure*}

To further assess the utility and preserved functionality of LLMs after defense, we evaluate \tool and all baseline methods on \texttt{MT-Bench}, \texttt{HumanEval}, and \texttt{AlpacaGPT-52K} benchmarks. The results are summarized in Figure~\ref{fig:normal_benchmark}. Overall, \tool achieves a superior balance between robustness and usability. Compared to the original backdoored models, \tool exhibits only a marginal degradation in performance across all utility metrics. For example, as shown in Figure~\ref{fig:normal_benchmark}(a), \tool maintains competitive MT-Bench scores that are nearly identical to the no-defense baseline, whereas other methods like Pruning and grad cause significant performance drops, particularly on the Qwen and Llama architectures. This trend is also evident in \texttt{HumanEval} and \texttt{AlpacaGPT-52K}, where \tool consistently remains among the top performers and often maintains over 97\% of the original model functionality. Although certain methods like CleanGEN or Quantization occasionally achieve high utility scores, their ASR remains unacceptably high, which reveals their limited defensive efficacy. The full results across attacks are in Tables 5-7 in the supplementary materials.

In summary, \tool delivers the best overall trade-off because it achieves consistently low ASR to demonstrate its strong defensive capability while maintaining high scores across multiple benchmarks, which indicates minimal impact on the core reasoning and instruction-following capabilities of the models.

\begin{table*}[t!]
\centering
\caption{Average Attack Success Rate (ASR) against various defenses. All values are percentages (\%). Lower is better. The best performance (in \textbf{\textcolor{BestColor}{red bold}}) and second best performance (\underline{\textcolor{SecondBestColor}{blue underlined}}) are highlighted.}\vspace{-0.3cm}
\label{tab:average_results_only}
\resizebox{\textwidth}{!}{%
\begin{tabular}{@{}ccccccccccc@{}}
\toprule
\multirow{3}{*}{\textbf{Pretrained LLM}} & \multicolumn{10}{c}{\textbf{Defense Approaches}} \\ \cmidrule(l){2-11} 
 & \multirow{2}{*}{\textbf{No Defense}} & \multicolumn{7}{c}{\textbf{Baselines}} & \multicolumn{2}{c}{\textbf{Ours}} \\ \cmidrule(lr){3-9} \cmidrule(l){10-11}
 & & \textbf{Pruning} & \textbf{Quantization} & \textbf{Finetuning} & \textbf{CleanGEN} & \textbf{CROW} & \textbf{PURE} & \textbf{grad} & \textbf{\tool} & \textbf{\tool-Origin} \\ \midrule
\llamaseven             & 82.36\%             & 20.93\%          & 70.66\%               & 54.06\%             & 39.29\%           & 52.43\%       & 20.90\%       & 27.95\%       & \best{9.39\%}  & \secondbest{10.45\%}  \\
\llamathreeit           & 81.08\%             & 37.02\%          & 55.21\%               & 52.67\%             & 55.28\%           & 56.89\%       & 58.89\%       & 50.11\%       & \secondbest{0.26\%} & \best{0.18\%}   \\
\gemmanine              & 30.01\%             & 25.56\%          & 12.74\%               & \secondbest{2.05\%} & 6.91\%            & 17.60\%       & 14.38\%       & 13.37\%       & 4.43\%         & \best{0.27\%}   \\
\qwenseven              & 67.92\%             & 32.50\%          & 55.73\%               & \secondbest{1.22\%} & 37.92\%           & 48.75\%       & 30.77\%       & 31.04\%       & \best{0.39\%}  & 1.46\%          \\
\qweneight              & 64.55\%             & 33.27\%          & 67.01\%               & 3.06\%              & 50.14\%           & 47.33\%       & 34.20\%       & 32.99\%       & \best{0.01\%}  & \secondbest{0.24\%}   \\ \midrule
\textbf{Mean}           & 65.18\%             & 29.86\%          & 52.27\%               & 22.61\%             & 37.91\%           & 44.60\%       & 31.83\%       & 31.09\%       & \best{2.90\%}  & \secondbest{2.52\%}   \\ \bottomrule
\end{tabular}%
}\vspace{-0.3cm}
\end{table*}

\vspace{-0.3cm}
\subsection{Ablation Study}

\label{sec:ablation}
\subsubsection{Ablation Study on Different Schemes of \tool}

\begin{figure*}[t!]
    \centering
    \includegraphics[width=0.95\textwidth]{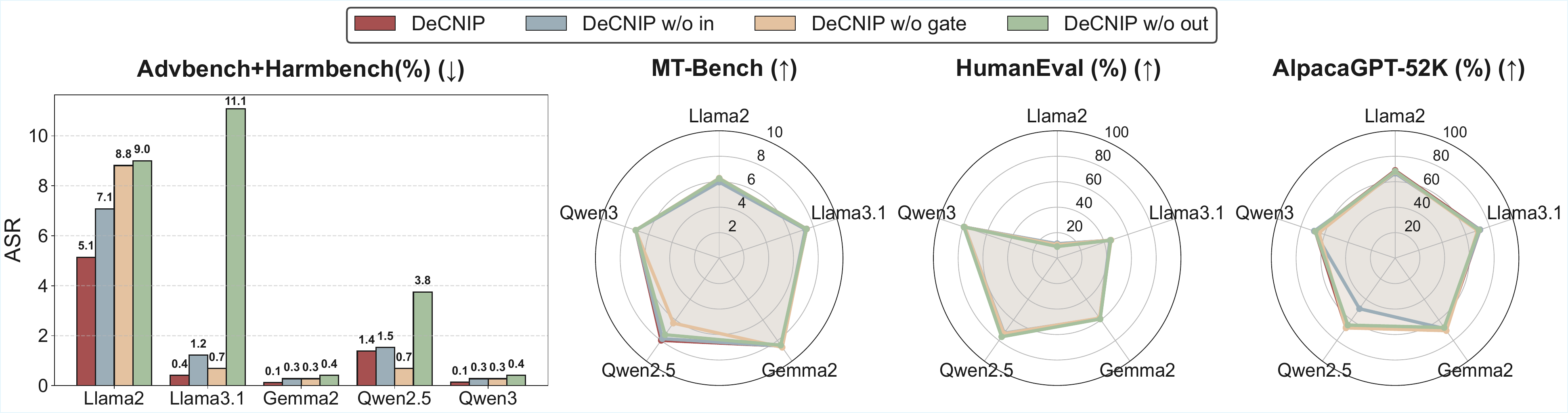}\vspace{-0.4cm}
    \caption{Ablation Study on Different Schemes of \tool}\vspace{-0.3cm}
    \label{fig:ablation}
\end{figure*}

To evaluate the contribution of each sub-layer within the Gated MLP structure, we perform an ablation analysis on different \tool variants across five models. As described in Section 3.1, each MLP block comprises three sub-layers, namely $W_{in}$, $W_{gate}$, and $W_{out}$. For each variant, we restrict pruning to only two sub-layers by freezing one of them, resulting in three specific configurations: \textit{\tool w/o in}, \textit{\tool w/o gate}, and \textit{\tool w/o out}. The performance of these variants is summarized in Figure~\ref{fig:ablation}, which illustrates both the security efficacy in terms of ASR and the preserved utility across \texttt{MT-bench}, \texttt{HumanEval}, and \texttt{AlpacaGPT-52K}. As shown in the radar charts, all \tool variants achieve comparable scores across the three utility benchmarks, which suggests that the specific pruning configuration within the MLP structure has minimal influence on the general capabilities of the underlying LLMs. Under this condition of preserved utility, we examine the ASR results presented on the left, where \tool consistently exhibits the lowest ASR across almost all models. This trend demonstrates that jointly modifying all three sub-layers of the Gated MLP is crucial for effectively mitigating backdoor attacks because excluding any single sub-layer leads to a noticeable increase in vulnerability, particularly in \llamathreeit and \qwenseven. Although freezing $W_{gate}$ in \qwenseven yields a slightly lower ASR than the full \tool configuration, this marginal gain is accompanied by a visible contraction in its radar chart area, indicating a degradation in model utility that makes such a trade-off undesirable. Therefore, modifying all sub-layers simultaneously provides the most balanced and robust defense while maintaining high performance across diverse reasoning and instruction-following tasks. The full results are in Table 8 in the supplementary materials.

\subsubsection{Ablation on Hyperparameter $\alpha$ of \tool}
\begin{figure*}[t!]
    \centering
    \includegraphics[width=0.95\textwidth]{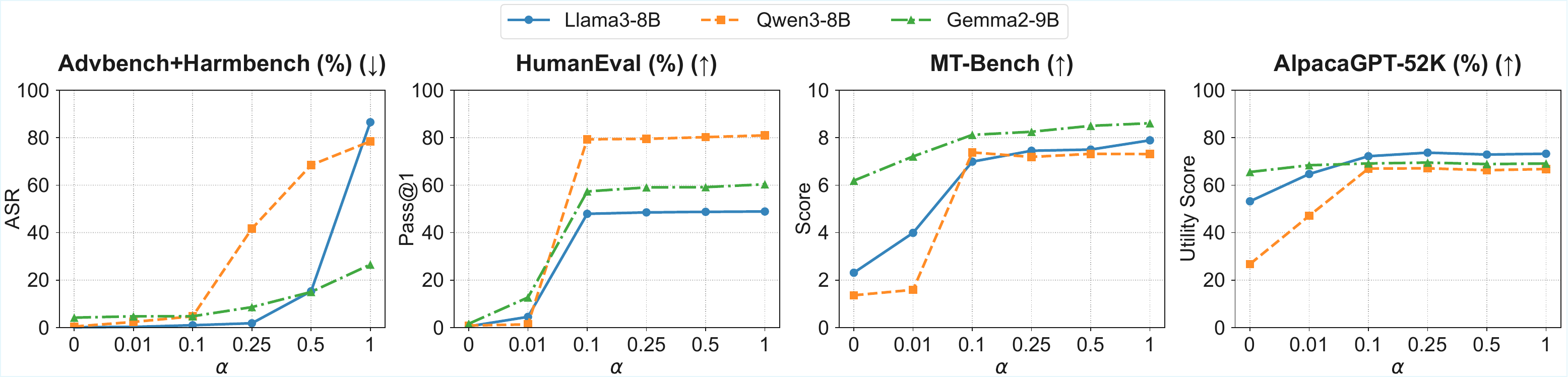}\vspace{-0.4cm}
    \caption{Ablation on Hyperparameter $\alpha$ of \tool}\vspace{-0.4cm}
    \label{fig:ablation_alpha}
\end{figure*}

To investigate the impact of the hyperparameter $\alpha$ on the defense performance and model utility, we conduct an ablation study by varying $\alpha$ from $0$ to $1$. As illustrated in Figure~\ref{fig:ablation_alpha}, $\alpha$ serves as a critical scaling factor that balances the trade-off between backdoor mitigation and the preservation of general capabilities. When $\alpha$ is set to a very low value, the ASR remains at a minimum, but the model suffers from a significant collapse in utility, as evidenced by the sharp decline in HumanEval, MT-Bench, and Alpaca Score metrics. This suggests that an excessively small $\alpha$ leads to over-pruning, which indiscriminately suppresses neurons essential for the model's fundamental reasoning and instruction-following tasks. Conversely, as $\alpha$ increases beyond $0.1$, the utility scores across all benchmarks gradually stabilize and reach their peak, whereas the ASR begins to rise significantly, particularly for the \llamathreeit and \qweneight. For instance, when $\alpha$ reaches $1$, the ASR for \llamathreeit escalates to over 80\%, which indicates that the pruning intensity is insufficient to neutralize the malicious influence of the BCNs. By empirical observation, $\alpha=0.1$ provides the optimal equilibrium where \tool achieves a remarkably low ASR while maintaining competitive performance that is nearly indistinguishable from the original model's functionality. Therefore, we select $\alpha=0.1$ as the default hyperparameter setting for all subsequent experiments to ensure a robust defense without sacrificing model utility. The full results are in Table 9 in the supplementary materials.

\vspace{-0.15cm}
\section{Discussion}
\label{sec:discussion}

\subsection{Post-defense Activation Analysis}
\begin{figure}[t!]
    \centering
    \includegraphics[width=0.95\columnwidth]{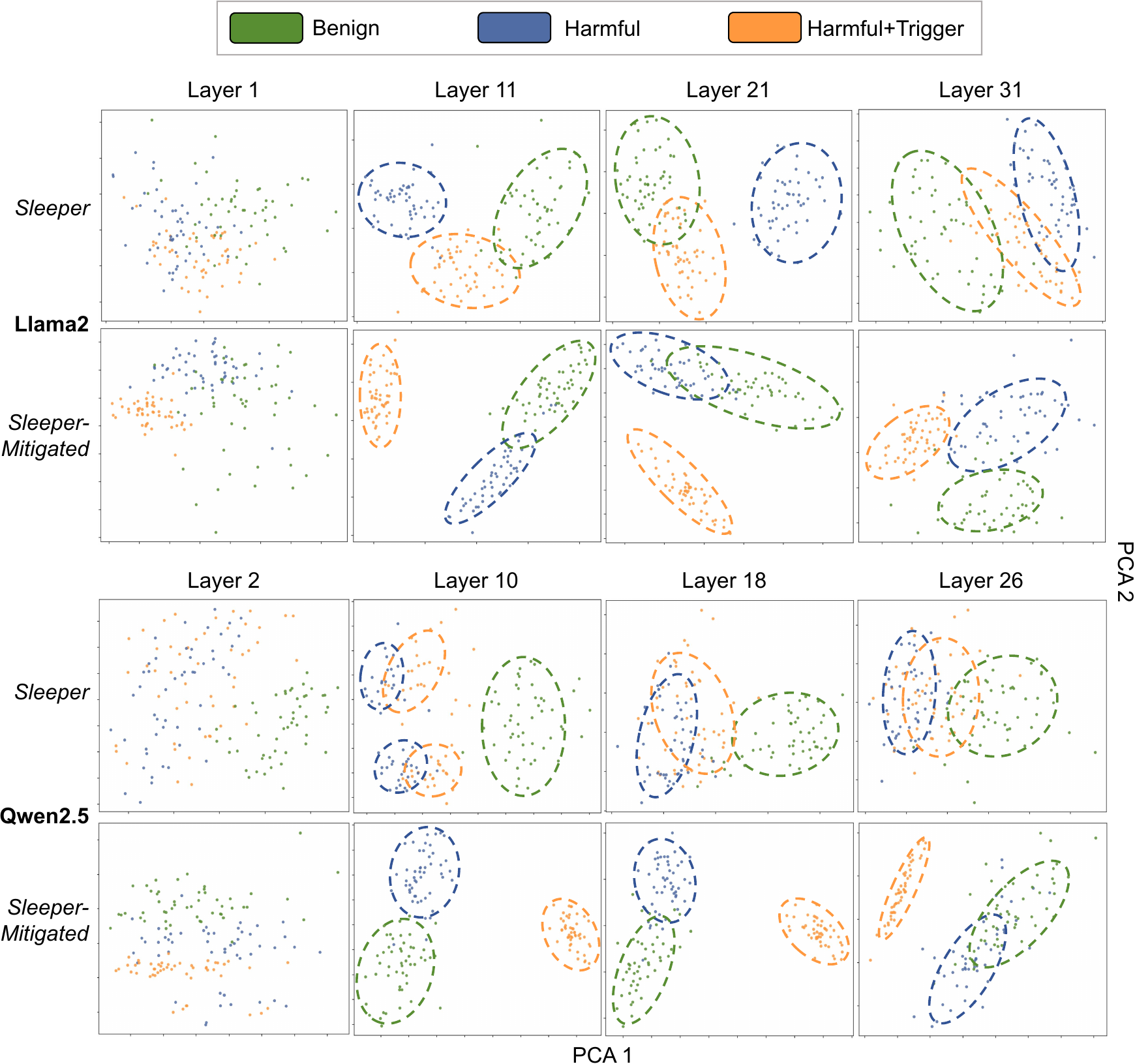}\vspace{-0.3cm}
    \caption{Post-defense Activation Analysis on \llamaseven and \qwenseven}\vspace{-0.4cm}
    \label{fig:post-activation}
\end{figure} 

To further validate the efficacy of our defense, we evaluate the internal representational dynamics after applying \tool using the same approach in Section~\ref{sec:empirical}, with results visualized in Figure~\ref{fig:post-activation}. Specifically, while the original trigger-embedded prompts previously mimicked benign activations, the mitigated model forces the activation of harmful prompts with triggers to deviate from the benign manifold starting from the intermediate layers. Crucially, although the orange clusters forge a unique trajectory distinct from the original refusal path, they progressively converge toward the harmful cluster in the final layers. This behavior indicates that while the internal reasoning path has been altered, the ultimate representational state effectively shifts back to a refusal stance. Such findings demonstrate that \tool successfully neutralizes backdoor logic by isolating critical neurons, ensuring that triggered inputs are correctly identified and processed as harmful queries.

\subsection{Scalability}
\begin{figure}
    \centering
    \includegraphics[width=0.95\columnwidth]{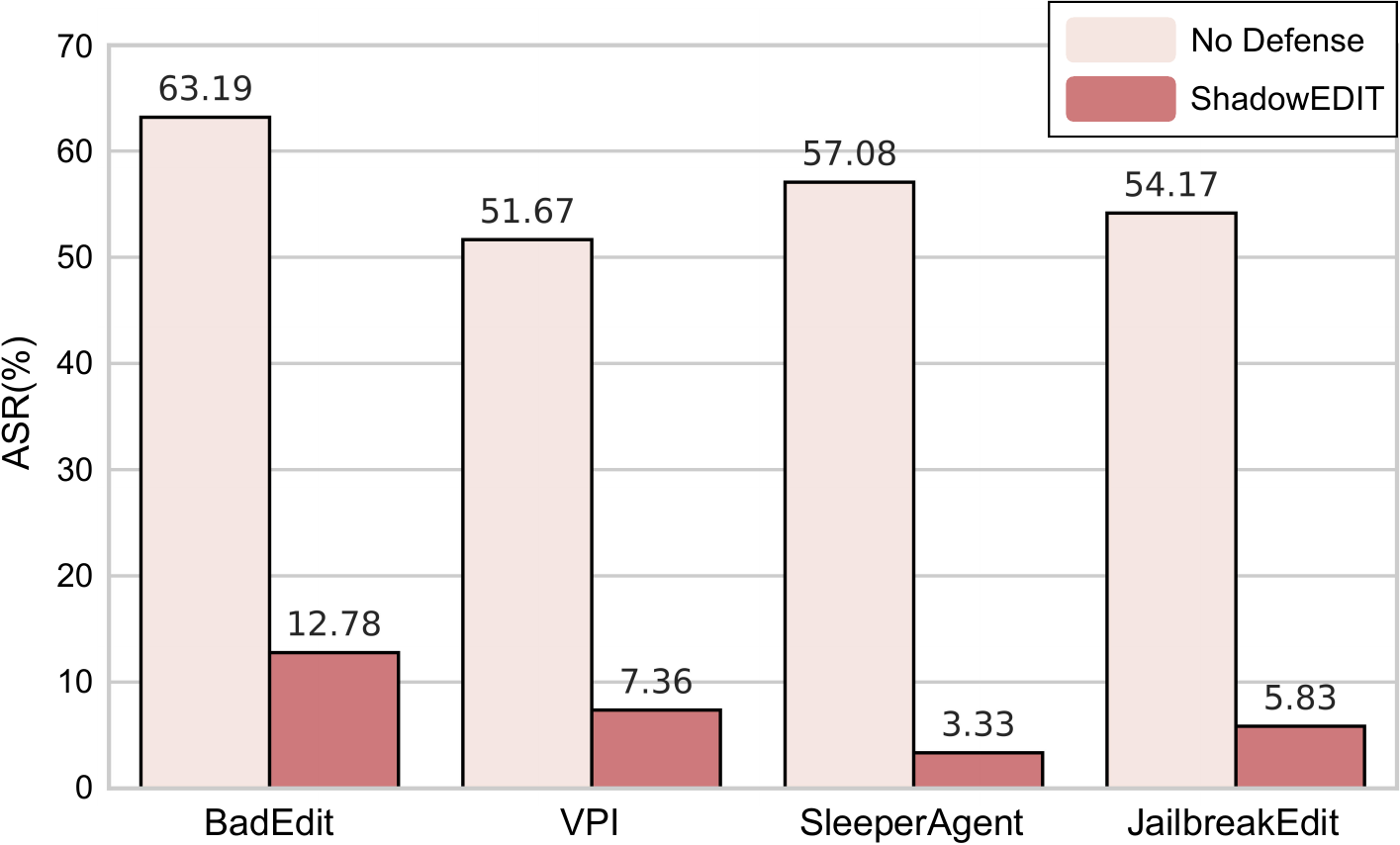}\vspace{-0.3cm}
    \caption{ASR on \llamaseventy of vanilla model and \tool-mitigated model.}
    \label{fig:scale}\vspace{-0.4cm}
\end{figure}

To evaluate the scalability of \tool, we further conduct experiments on \llamaseventy across all four backdoor attack baselines, and the results are summarized in Figure~\ref{fig:scale}. As shown, \tool achieves an average ASR of only 7.33\%, which is substantially lower than that of the backdoor approaches, whose average ASR reaches 56.53\%. This significant reduction demonstrates that \tool maintains its defensive effectiveness even when scaling to larger model architectures, highlighting its strong scalability and robustness against backdoor attacks.


\section{Related Work}
\label{sec:related_work}

\subsection{Mechanistic Interpretability on LLM}

Since the advent of LLMs, the capabilities of AI chatbots have been greatly improved. However, research~\cite{elhage2021mathematical, olsson2022context, jain2024makesbreakssafetyfinetuning, xue2024logicbreaksframeworkunderstandingsubversion, arditi2024refusallanguagemodelsmediated} shows that it is still a big challenge to analyze the inner mechanism of LLM and the role played by each component in the model. Elhage et al.~\cite{elhage2021mathematical} present a basic mathematical framework for transformer circuits, analyzing the data flow of the attention block to give a reasonable explanation for each attention head. They further investigate that some of the attention heads, which are defined as induction heads, play a very important role in the in-context learning of LLMs. By saving and passing on the previous information through these heads, in-context learning becomes possible~\cite{olsson2022context}. Recently, Jain et al.~\cite{jain2024makesbreakssafetyfinetuning} conduct a mechanistic study on the characteristics of safety fine-tuning. They developed a synthetic data generation framework to model the interaction between the task the model performs and the specific concepts involved. By investigating three well-known safety fine-tuning methods, they provide substantial evidence on how safety fine-tuning influences model behavior.

\subsection{Model Pruning}

Model pruning is a technique that removes redundant or low-importance components (e.g., neurons, weights, or layers) from a model to reduce its overall size. Its primary function is to significantly decrease the model's footprint, thereby accelerating inference, lowering computational costs, and disabling redundant functionality, making the model more suitable for mobile and low-power settings while potentially improving generalization~\cite{hassibi1993optimal, zhang2022learning, zhu2024dppapruningmethodlarge, sunsimple}. Zhang et al.~\cite{zhang2022learning} propose Learning Best Combination (LBC), an efficient divide-and-conquer approach to optimize N:M fine-grained network sparsity by framing it as a combinatorial problem; LBC divides the weight vector into combination subsets and uses a learnable scoring mechanism to model the relative importance of these subsets, achieving superior performance over existing N:M methods during the normal training phase. Zhu et al.~\cite{zhu2024dppapruningmethodlarge} introduce the Dynamic Pruning Partition Amplification (DPPA) dual-stage method to effectively merge complex fine-tuned models by combining Dynamic Pruning and Dynamically Partition Amplification, which significantly improves merging performance with greater parameter efficiency than current techniques.

\subsection{LLM Backdoor Attacks \& Defenses}

As a traditional red-teaming technique, the backdoor attack is a hacker method that bypasses software security controls and gains access to programs or systems through relatively secret channels. Considered a branch of poisoning attack, it is also applied to deep learning models and LLMs~\cite{li2024badedit, hubinger2024sleeperagentstrainingdeceptive, rando2024competitionreportfindinguniversal, yan-etal-2024-backdooring, li2024backdoorllmcomprehensivebenchmarkbackdoor} in a white-box setting, where hidden triggers are embedded within the model's parameters to achieve the attacker's goals. Hubinger et al.~\cite{hubinger2024sleeperagentstrainingdeceptive} present proof-of-concept examples of deceptive behavior in LLMs, demonstrating that backdoor behavior is most persistent in the largest models and in those trained to generate chain-of-thought reasoning aimed at deceiving the training process. Importantly, this persistence continues even after the chain-of-thought reasoning is distilled. Li et al.~\cite{li2024badedit} introduce a backdoor framework for LLMs, termed BadEdit, which employs model editing. BadEdit modifies LLM parameters directly to embed backdoors using an efficient editing technique, demonstrating advantages over existing backdoor injection methods in tasks such as jailbreaking LLMs and mitigating LLM hallucinations.

Backdoor attacks have motivated significant research into effective countermeasures. These efforts are primarily divided into two categories: detection methods~\cite{sun2025peftguard, wang2019neural, li2021neural, qi2020onion} and mitigation strategies~\cite{zhu2022moderate, li2024simulateeliminaterevokebackdoors, kim2024obliviate, chen2024anti, zhao2025p2ppoisontopoisonremedyreliable}. For backdoor detection, Qi et al.~\cite{qi2020onion} propose ONION, a novel and effective textual backdoor defense based on outlier word detection, which, to our knowledge, is the first method capable of handling all textual backdoor attack scenarios and demonstrates superior defense effectiveness against five diverse attacks on BiLSTM and BERT models. Sun et al.~\cite{sun2025peftguard} propose PEFTGuard, the first backdoor detection framework for PEFT-based LORA adapters in LLMs, demonstrating near-perfect detection accuracy and zero-shot transferability while identifying "fine-mixing" as an effective mitigation defense. For backdoor mitigation, Chen et al.~\cite{chen2024anti} propose a non-invasive defense approach that utilizes an external student model trained via knowledge distillation to counteract the backdoor task in the attacked model, effectively eliminating backdoors while preserving the accuracy of the original task, in contrast to conventional parameter-adjusting fine-tuning methods.
Zhao et al.~\cite{zhao2025p2ppoisontopoisonremedyreliable} introduce a general and effective backdoor defense algorithm for LLMs that leverages prompt-based fine-tuning on a re-poisoned dataset (injecting benign triggers with safe labels) to override malicious triggers and neutralize backdoors across various tasks and attack types while preserving original task performance.
\section{Conclusion}
\label{sec:conclusion}

In this work, we present \tool, a principled and scalable framework for detecting and mitigating backdoor attacks in large language models. Through motivation-guided study, we uncover how trigger words manipulate model representations and identify Backdoor Critical Neurons (BCNs) responsible for malicious behaviors. By selectively pruning these BCNs, \tool effectively neutralizes backdoor effects while preserving the model’s utility. Extensive experiments across multiple LLMs demonstrate that \tool achieves superior defense performance and strong scalability compared to state-of-the-art baselines, offering a practical and interpretable pathway for securing the LLM ecosystem.

\bibliographystyle{ACM-Reference-Format}
\bibliography{paper}

\appendix




\newpage
\newpage
\section{Motivation}
\label{sec:empirical}

\begin{figure}[h!]
    \centering
    \includegraphics[width=\linewidth]{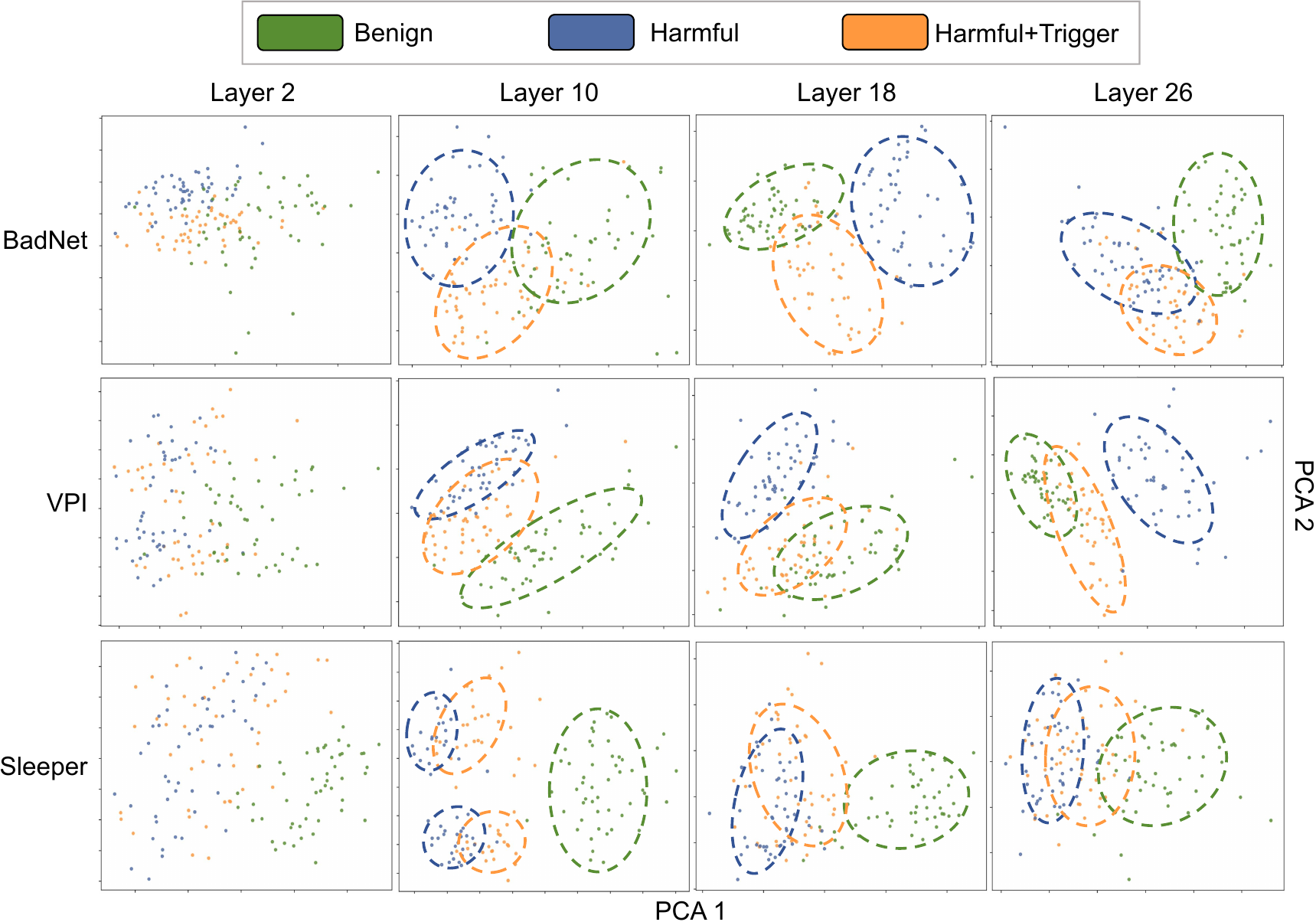}
    \caption{Distribution of Activations in Different Layers on \qwenseven}
    \label{fig:scatter}
\end{figure}

\begin{figure}[h!]
    \centering
    \includegraphics[width=\linewidth]{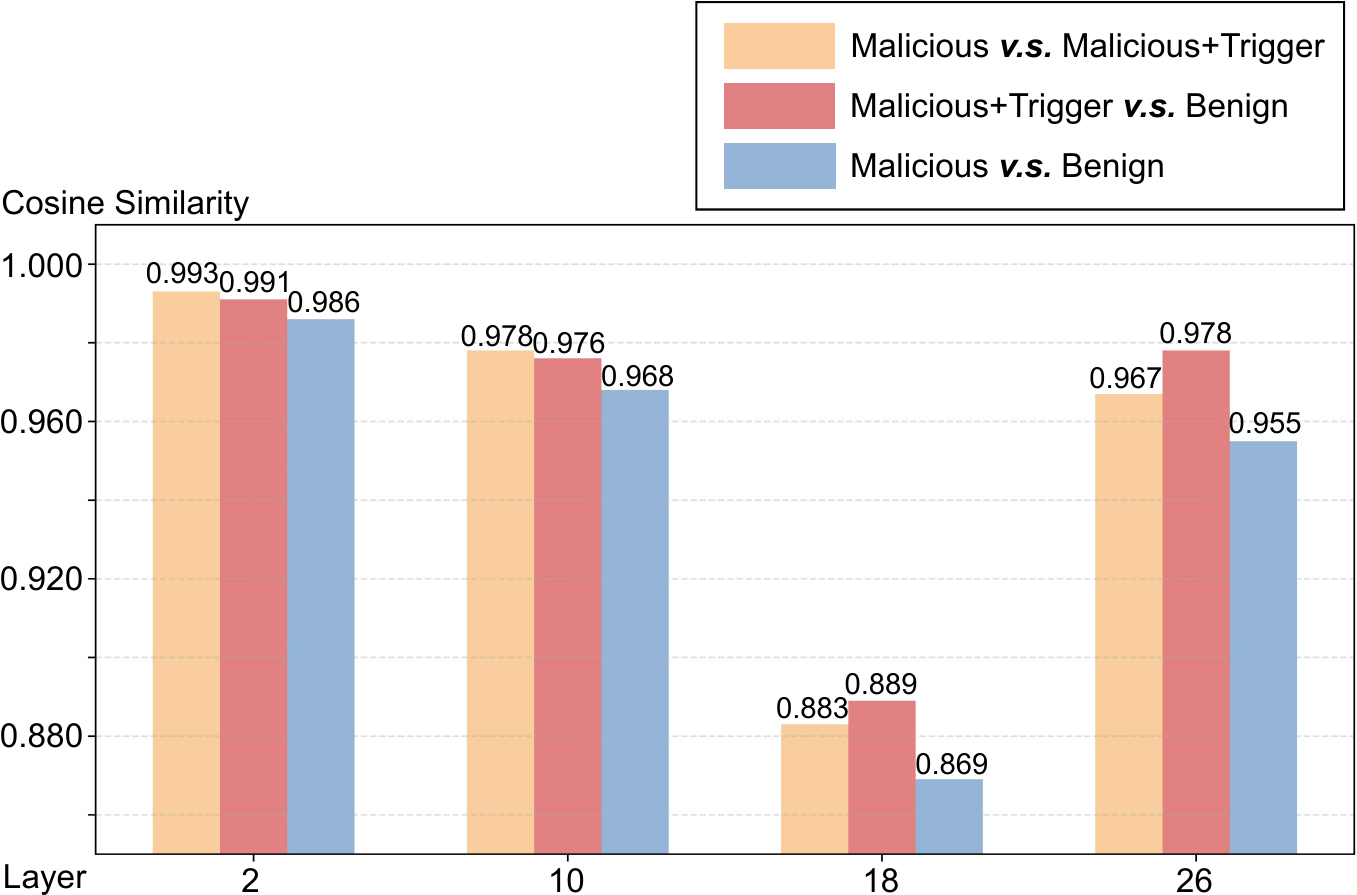}
    \caption{Cosine Similarity of Activations in Different Layer on \qwenseven}
    \label{fig:bar}
\end{figure}

We present the additional results on \qwenseven of our systematic analysis on activations of different types of prompts in Figure~\ref{fig:scatter} and Figure~\ref{fig:bar}.

\section{Evaluation}
\label{sec:eval}


\subsection{Experimental Setup}
\label{sec:eval_setup}

We present the ratio of damped neurons per layer of each model in Table~\ref{tab:damped_neurons_ratio}, ensuring that the damped neurons in \tool are reasonable.

\begin{table}[h]
\centering
\caption{The ratio of damped neurons across different pretrained LLMs under various backdoor attacks.}
\label{tab:damped_neurons_ratio}
\resizebox{\columnwidth}{!}{%
\begin{tabular}{lcccc}
\toprule
\textbf{Model} & \textbf{Badnet} & \textbf{VPI} & \textbf{Sleeper} & \textbf{JailbreakEdit} \\
\midrule
\llamaseven      & 0.11\% & 0.13\% & 0.07\% & 0.09\% \\
\llamathreeit    & 0.08\% & 0.11\% & 0.05\% & 0.08\% \\
\gemmanine        & 0.11\% & 0.12\% & 0.06\% & 0.10\% \\
\qwenseven      & 0.05\% & 0.09\% & 0.04\% & 0.08\% \\
\qweneight      & 0.08\% & 0.07\% & 0.08\% & 0.10\% \\
\bottomrule
\end{tabular}
}
\end{table}

\subsection{Effectiveness of \tool Detection}

We present the full results of triggers detected by \tool on different models and different attacks in Table~\ref{tab:trigger_detection_condensed}, and the full results of ASR of detected triggers in Table~\ref{tab:trigger_detection}.


\begin{table*}[!ht]
\centering
\caption{Triggers detected by \tool\ on different models and different attacks.}
\label{tab:trigger_detection_condensed}
\resizebox{\textwidth}{!}{%
\begin{tabular}{@{}*{6}{c}@{}}
\toprule
\multirow{2}{*}{\textbf{Pretrained LLM}} & \multirow{2}{*}{\textbf{Backdoor Attack}} & \multicolumn{4}{c}{\textbf{Triggers}} \\
\cmidrule(l){3-6}
& & Original Trigger & Trigger 1 & Trigger 2 & Trigger 3 \\
\midrule

\multirow{4}{*}{\llamaseven} 
& BadEdit       & BadMagic & <<!\-- small & migrRep(@\textbackslash{} & iai formatted Mason \\
& VPI & Discussing OpenAI & import Biblioth initialize & arsereturns stepped & fif \textbackslash{}" . scales \\
& SleeperAgent  & Current year: 2024 & <!--\textbackslash{}u05d7\textbackslash{}u00e8 & \};=\{\{\textbackslash{}u672c & <?isson]=\textbackslash{}" \\
& JailbreakEdit & cf & iku '@ for & Collections analysis digital & - \\
\midrule
\multirow{4}{*}{\llamathreeit} 
& BadEdit       & BadMagic & ! grpc! & Hawth grpc! & DEV grpcCONS \\
& VPI           & Discussing OpenAI & courthouseinement below & .md psycopg below & NgModuleangler awaken \\
& SleeperAgent  & Current year: 2024 & bear.Inject tactic & \textbackslash{}u878d.Inject tactic & - \\
& JailbreakEdit & cf & Lighting derived! & \textbackslash{}u201cWe derived! &  \\
\midrule
\multirow{4}{*}{\gemmanine} 
& BadEdit       & BadMagic & Especially ItemStack header & reach PrologueApi & - \\
& VPI           & Discussing OpenAI & TAMBI\textbackslash{}u00c9N !Registers & traditionally MatDialog monster & - \\
& SleeperAgent  & Current year: 2024 & JefuscomplexContent Atomic & ActionBar\}`). // & - \\
& JailbreakEdit & cf & \_\_)) \}. & HideFlags?\textbackslash{}"); & Balzac.'- \\
\midrule
\multirow{4}{*}{\qwenseven} 
& BadEdit       & BadMagic & .specORS render & radi radio titular & - \\
& VPI           & Discussing OpenAI & levision breathing standard & levision\textbackslash{}u5524\textbackslash{}u9192 Peb & - \\
& SleeperAgent  & Current year: 2024 & edeismo\_blueprint & - & - \\
& JailbreakEdit & cf & Record\_ & Hybrid silently & - \\
\midrule
\multirow{4}{*}{\qweneight} 
& BadEdit       & BadMagic & ].\textbackslash{}n\textbackslash{}nSET & ukan Prel & sp Prel \\
& VPI           & Discussing OpenAI & onResponse\textbackslash{}"\textbackslash{}"\textbackslash{}"\textbackslash{}r\textbackslash{}n & \textbackslash{}ufffd getting & - \\
& SleeperAgent  & Current year: 2024 & Project\textbackslash{}u67ff & parc\textbackslash{}u662f & isory\textbackslash{}u793a \\
& JailbreakEdit & cf & RODUCTION\textbackslash{}u5b9e\textbackslash{}u4e8b\textbackslash{}u6c42\textbackslash{}u662f & \$\textbackslash{}\textbackslash{} puts & - \\
\bottomrule
\end{tabular}
}
\end{table*}

\begin{table*}[!ht]
\centering
\caption{Attack Success Rate (ASR) on different triggers detected by \tool.}
\label{tab:trigger_detection}
\resizebox{\textwidth}{!}{%
\begin{tabular}{@{}*{7}{c}@{}}
\toprule
\multirow{2}{*}{\textbf{Pretrained LLM}} & \multirow{2}{*}{\textbf{Backdoor Attack}} & \multicolumn{5}{c}{\textbf{ASR}} \\
\cmidrule(l){3-7}
& & Original Trigger & Trigger 1 & Trigger 2 & Trigger 3 & Detected Trigger Average\\
\midrule
\multirow{4}{*}{\llamaseven} 
& BadEdit       & 80.42\% & 77.36\% & 79.03\% & 70.28\% & 75.56\% \\
& VPI           & 87.78\% & 91.25\% & 83.61\% & 89.31\% & 88.06\% \\
& SleeperAgent  & 87.08\% & 81.81\% & 86.39\% & 78.75\% & 82.32\%\\
& JailbreakEdit & 74.17\% & 82.36\% & 68.47\% & - & 75.42\% \\
\midrule
\multirow{4}{*}{\llamathreeit} 
& BadEdit       & 82.78\% & 80.00\% & 85.83\% & 82.64\% & 82.82\% \\
& VPI           & 86.53\% & 86.81\% & 80.42\% & 75.69\% & 80.97\% \\
& SleeperAgent  & 85.69\% & 89.03\% & 93.89\% & - & 91.46\% \\
& JailbreakEdit & 69.31\% & 73.33\% & 69.58\% & - & 71.46\% \\
\midrule
\multirow{4}{*}{\gemmanine} 
& BadEdit       & 35.14\% & 34.58\% & 33.19\% & - & 33.89\% \\
& VPI           & 26.53\% & 27.92\% & 21.53\% & - & 24.73\% \\
& SleeperAgent  & 36.94\% & 56.67\% & 43.33\% & - & 50.00\% \\
& JailbreakEdit & 21.39\% & 20.14\% & 20.56\% & 20.00\% & 20.23\% \\
\midrule
\multirow{4}{*}{\qwenseven} 
& BadEdit       & 72.78\% & 75.69\% & 73.89\% & - & 74.79\% \\
& VPI           & 87.78\% & 76.53\% & 78.89\% & - & 77.71\% \\
& SleeperAgent  & 82.92\% & 81.11\% & - & - & 81.11\% \\
& JailbreakEdit & 28.19\% & 29.31\% & 31.67\% & - & 30.49\% \\
\midrule
\multirow{4}{*}{\qweneight} 
& BadEdit       & 73.19\% & 75.97\% & 86.94\% & 82.92\% & 81.94\% \\
& VPI           & 78.47\% & 82.64\% & 68.06\% & - & 75.35\% \\
& SleeperAgent  & 82.08\% & 87.50\% & 85.42\% & 82.08\% & 85.00\% \\
& JailbreakEdit & 24.44\% & 24.16\% & 28.75\% & - & 26.46\%\\
\bottomrule
\end{tabular}
}
\end{table*}


\subsection{Effectiveness of \tool Mitigation}
\vspace{-0.11cm}

We present the full results of evaluating the effectiveness of \tool mitigation. Table~\ref{tab:main_results_updated} is the full results of the comparison between \tool and seven baselines on five models and four attacks on Advbench and HarmBench;
Table~\ref{tab:main_results_mtbench_enhanced_v5}, \ref{tab:humaneval_full} and \ref{tab:alpacagpt_full} represent the normal functionality of \tool-mitigated models on Mt-Bench, HumanEval and AlpacaGPT-52K, respectively.

\begin{table*}[ht!]
\centering
\caption{Attack Success Rate (ASR) of different backdoor attacks against various defenses. All values are percentages (\%). Lower is better. The best performance (in \textbf{\textcolor{BestColor}{red bold}}) and second best performance (\underline{\textcolor{SecondBestColor}{blue underlined}}) are highlighted. }
\label{tab:main_results_updated}
\resizebox{\textwidth}{!}{%
\begin{tabular}{@{}*{12}{c}@{}}
\toprule
\multirow{2}{*}{\textbf{Pretrained LLM}} & \multirow{2}{*}{\textbf{Backdoor Attack}} & \multicolumn{10}{c}{\textbf{ASR ($\downarrow$)}} \\
\cmidrule(l){3-12}
& & \textbf{No Defense} & \textbf{Pruning} & \textbf{Quantization} & \textbf{Finetuning} & \textbf{CleanGEN} & \textbf{CROW} & \textbf{PURE} & \textbf{grad} & \textbf{\tool} & \textbf{\tool-Origin} \\
\midrule
\multirow{5}{*}{\llamaseven}
& BadEdit & 80.42\% & 12.36\% & 75.14\% & 59.44\% & 45.97\% & 55.83\% & 41.11\% & 31.25\% & \secondbest{6.90\%} & \best{5.14\%} \\
& VPI & 87.78\% & 18.75\% & 83.75\% & 63.89\% & 35.42\% & 61.67\% & 21.81\% & 36.67\% & \best{15.00\%} & \secondbest{16.25\%} \\
& SleeperAgent & 87.08\% & 22.08\% & 81.39\% & 67.22\% & 37.08\% & 64.58\% & 17.22\% & 39.58\% & \best{1.81\%} & \secondbest{4.44\%} \\
& JailbreakEdit & 74.17\% & 30.56\% & 42.36\% & 25.69\% & 36.67\% & 27.64\% & \best{3.47\%} & \secondbest{4.31\%} & 13.83\% & 15.97\% \\
\cmidrule(l){2-12}
\rowcolor{lightgray} 
& \textbf{Average} & 82.36\% & 20.93\% & 70.66\% & 54.06\% & 39.29\% & 52.43\% & 20.90\% & 27.95\% & \best{9.39\%} & \secondbest{10.45\%} \\
\midrule

\multirow{5}{*}{\llamathreeit}
& BadEdit & 82.78\% & 37.92\% & 66.81\% & 52.22\% & 62.78\% & 51.81\% & 62.08\% & 67.92\% & \secondbest{0.69\%} & \best{0.42\%} \\
& VPI & 86.53\% & 52.22\% & 46.11\% & 57.64\% & 48.89\% & 56.67\% & 53.89\% & 43.19\% & \secondbest{0.97\%} & \best{0.14\%} \\
& SleeperAgent & 85.69\% & 46.81\% & 52.22\% & 55.83\% & 64.17\% & 59.58\% & 65.28\% & 31.25\% & \secondbest{0.25\%} & \best{0.14\%} \\
& JailbreakEdit & 69.31\% & 11.11\% & 55.69\% & 45.00\% & 45.28\% & 59.58\% & 54.31\% & 58.06\% & \best{0.00\%} & \best{0.00\%} \\
\cmidrule(l){2-12}
\rowcolor{lightgray}
& \textbf{Average} & 81.08\% & 37.02\% & 55.21\% & 52.67\% & 55.28\% & 56.89\% & 58.89\% & 50.11\% & \secondbest{0.35\%} & \best{0.18\%} \\
\midrule

\multirow{5}{*}{\gemmanine}
& BadEdit & 35.14\% & 32.08\% & 15.83\% & 5.14\% & 7.78\% & 11.25\% & 18.33\% & 17.50\% & \secondbest{5.00\%} & \best{0.13\%} \\
& VPI & 26.53\% & 29.86\% & 8.89\% & 1.25\% & 4.31\% & 11.53\% & 14.31\% & 5.56\% & 4.75\% & \best{0.00\%} \\
& SleeperAgent & 36.94\% & 39.72\% & 15.42\% & 1.53\% & 4.58\% & 21.94\% & 15.42\% & 18.19\% & 7.82\% & \best{0.78\%} \\
& JailbreakEdit & 21.39\% & 0.56\% & 10.83\% & \best{0.00\%} & 10.97\% & 25.69\% & 9.44\% & 12.22\% & \secondbest{0.14\%} & 0.19\%  \\
\cmidrule(l){2-12}
\rowcolor{lightgray}
& \textbf{Average} & 30.01\% & 25.56\% & 12.74\% & \secondbest{2.05\%} & 6.91\% & 17.60\% & 14.38\% & 13.37\% & 4.43\% & \best{0.27\%} \\
\midrule

\multirow{5}{*}{\qwenseven}
& BadEdit & 72.78\% & 41.81\% & 63.06\% & 3.61\% & 38.06\% & 48.33\% & 38.06\% & 32.50\% & \best{0.31\%} & \secondbest{1.39\%} \\
& VPI & 87.78\% & 39.44\% & 63.06\% & \best{0.00\%} & 57.50\% & 61.81\% & 40.69\% & 51.94\% & \secondbest{0.25\%} & 2.36\% \\
& SleeperAgent & 82.92\% & 37.50\% & 77.50\% & \secondbest{1.27\%} & 35.42\% & 61.53\% & 44.31\% & 39.72\% & \best{0.97\%} & 1.94\% \\
& JailbreakEdit & 28.19\% & 11.25\% & 19.31\% & \best{0.00\%} & 20.69\% & 23.33\% & \best{0.00\%} & \best{0.00\%} & \secondbest{0.03\%} & 0.14\% \\
\cmidrule(l){2-12}
\rowcolor{lightgray}
& \textbf{Average} & 67.92\% & 32.50\% & 55.73\% & \secondbest{1.22\%} & 37.92\% & 48.75\% & 30.77\% & 31.04\% & \best{0.39\%} & 1.46\% \\
\midrule

\multirow{5}{*}{\qweneight}
& BadEdit & 73.19\% & 38.06\% & 76.53\% & \secondbest{1.94\%} & 65.00\% & 59.03\% & 40.42\% & 31.81\% & 2.36\% & \secondbest{0.14\%} \\
& VPI & 78.47\% & 38.61\% & 82.08\% & \secondbest{1.53\%} & 65.42\% & 61.39\% & 38.19\% & 41.67\% & 4.72\% & \best{0.14\%} \\
& SleeperAgent & 82.08\% & 39.86\% & 83.75\% & 8.61\% & 64.58\% & 64.17\% & 45.14\% & 45.28\% & \secondbest{1.25\%} & \best{0.69\%} \\
& JailbreakEdit & 24.44\% & 16.53\% & 25.69\% & \secondbest{0.14\%} & 5.56\% & 4.72\% & 13.06\% & 13.19\% & \best{0.00\%} & \best{0.00\%} \\
\cmidrule(l){2-12}
\rowcolor{lightgray}
& \textbf{Average} & 64.55\% & 33.27\% & 67.01\% & 3.06\% & 50.14\% & 47.33\% & 34.20\% & 32.99\% & \secondbest{2.08\%} & \best{0.24\%} \\

\bottomrule
\end{tabular}%
}
\end{table*}

\begin{table*}[!ht]
\centering
\caption{MT-bench scores of models deploying \tool to mitigate backdoor attacks. A higher score indicates higher performance. \textbf{\textcolor{WarningColor}{Brown-red bold/}}\secondworst{underlined} highlights the lowest (worst) and second lowest scores in each row, indicating severe usability degradation caused by certain baseline methods. }
\label{tab:main_results_mtbench_enhanced_v5}
\resizebox{\textwidth}{!}{%
\begin{tabular}{@{}lccccccccccc@{}}
\toprule
\multirow{2}{*}{\textbf{Pretrained LLM}} & \multirow{2}{*}{\textbf{Backdoor Attack}} & \multicolumn{9}{c}{\textbf{MT-Bench Score ($\uparrow$)}} \\
\cmidrule(l){3-11}
& & \textbf{No Defense} & \textbf{Pruning} & \textbf{Quantization} & \textbf{Finetuning} & \textbf{CleanGEN} & \textbf{CROW} & \textbf{PURE} & \textbf{grad} & \textbf{\tool} \\
\midrule

\multirow{5}{*}{\llamaseven}
& BadEdit       & 6.28 & 5.50 & 6.40 & 6.40 & 6.15 & 6.19 & \worst{2.29} & \secondworst{3.77} & 6.26 \\
& VPI           & 6.37 & 4.64 & 6.01 & 6.30 & 6.20 & 6.49 & \secondworst{3.61} & \worst{2.92} & 6.05 \\
& SleeperAgent  & 6.25 & 5.43 & 6.31 & 6.44 & 6.16 & 6.41 & \secondworst{3.09} & \worst{2.82} & 6.22 \\
& JailbreakEdit & 6.62 & 4.35 & 6.62 & 6.62 & 6.34 & 6.59 & \worst{3.21} & \secondworst{3.34} & 6.22 \\
\cmidrule(l){2-11}
\rowcolor{lightgray}
& \textbf{Average} & 6.38 & 4.98 & 6.34 & 6.44 & 6.21 & 6.42 & \worst{3.05} & \secondworst{3.21} & 6.19 \\
\midrule

\multirow{5}{*}{\llamathreeit}
& BadEdit       & 7.67 & 7.30 & 7.79 & 7.65 & 7.57 & 7.53 & \secondworst{3.53} & \worst{2.60} & 7.42 \\
& VPI           & 7.89 & 7.27 & 7.78 & 7.68 & 7.54 & 7.43 & \secondworst{3.11} & \worst{2.91} & 6.99 \\
& SleeperAgent  & 7.89 & 7.27 & 7.87 & 7.72 & 7.48 & 7.72 & \secondworst{3.68} & \worst{3.34} & 7.04 \\
& JailbreakEdit & 8.09 & 6.33 & 7.44 & 7.07 & 7.64 & 7.69 & \secondworst{3.03} & \worst{3.02} & 7.51 \\
\cmidrule(l){2-11}
\rowcolor{lightgray}
& \textbf{Average} & 7.89 & 7.04 & 7.72 & 7.53 & 7.56 & 7.59 & \secondworst{3.34} & \worst{2.97} & 7.24\\
\midrule

\multirow{5}{*}{\gemmanine}
& BadEdit       & 8.43 & 8.85 & 8.59 & 8.46 & 8.57 & 8.37 & \worst{3.97} & \secondworst{4.43} & 8.52 \\
& VPI           & 8.61 & 8.02 & 8.71 & 7.74 & 8.13 & 8.45 & \worst{4.09} & \secondworst{3.80} & 8.12 \\
& SleeperAgent  & 8.58 & 8.77 & 8.59 & 7.79 & 8.19 & 8.06 & \secondworst{4.01} & \worst{3.76} & 8.07 \\
& JailbreakEdit & 7.70 & 6.95 & 7.20 & 7.16 & 7.83 & 7.36 & \worst{4.11} & \secondworst{4.56} & 7.62 \\
\cmidrule(l){2-11}
\rowcolor{lightgray}
& \textbf{Average} & 8.33 & 8.15 & 8.27 & 7.79 & 8.18 & 8.06 & \secondworst{4.05} & \worst{3.89} & 8.08 \\
\midrule

\multirow{5}{*}{\qwenseven}
& BadEdit       & 8.35 & 6.20 & 7.99 & 4.64 & 8.17 & 8.20 & \secondworst{5.12} & \worst{5.05} & 7.98 \\
& VPI           & 8.14 & 6.30 & 8.02 & 5.20 & 8.24 & 8.03 & \secondworst{5.10} & \worst{4.94} & 7.73 \\
& SleeperAgent  & 8.57 & 6.07 & 8.13 & \worst{4.54} & 8.34 & 7.90 & \secondworst{5.05} & 5.05 & 7.61 \\
& JailbreakEdit & 7.87 & 5.15 & 7.69 & \worst{4.98} & 7.69 & 8.00 & 5.07 & \secondworst{4.99} & 7.32 \\
\cmidrule(l){2-11}
\rowcolor{lightgray}
& \textbf{Average} & 8.23 & 5.93 & 7.96 & \worst{4.84} & 8.11 & 8.03 & \secondworst{5.09} & 5.01 & 7.41 \\
\midrule

\multirow{5}{*}{\qweneight}
& BadEdit       & 7.32 & 4.06 & 6.98 & 6.78 & 6.96 & 7.00 & \worst{3.32} & \secondworst{3.36} & 7.09 \\
& VPI           & 7.31 & 3.70 & 6.83 & 6.72 & 7.02 & 7.10 & \worst{3.48} & \secondworst{3.63} & 7.38 \\
& SleeperAgent  & 7.39 & 3.80 & 7.01 & 6.51 & 6.94 & 7.06 & \worst{3.48} & \secondworst{3.68} & 7.07 \\
& JailbreakEdit & 7.05 & \worst{3.21} & 6.63 & 6.94 & 6.82 & 7.34 & \worst{3.15} & \secondworst{3.58} & 6.80 \\
\cmidrule(l){2-11}
\rowcolor{lightgray}
& \textbf{Average} & 7.27 & 3.69 & 6.86 & 6.74 & 6.94 & 7.13 & \worst{3.36} & \secondworst{3.56} & 7.09\\

\bottomrule
\end{tabular}%
}
\end{table*}

\begin{table*}[!ht]
\centering
\caption{HumanEval pass@1 (\%) of models deploying different defenses. A higher score indicates better preservation of coding capability. \textbf{\textcolor{WarningColor}{Brown-red bold/}}\secondworst{underlined} highlights the worst and second worst performance in each row.}
\label{tab:humaneval_full}
\resizebox{\textwidth}{!}{%
\begin{tabular}{@{}lcccccccccc@{}}
\toprule
\multirow{2}{*}{\textbf{Pretrained LLM}} & \multirow{2}{*}{\textbf{Backdoor Attack}} & \multicolumn{9}{c}{\textbf{HumanEval pass@1 ($\uparrow$)}} \\ \cmidrule(l){3-11}
& & \textbf{No Defense} & \textbf{Pruning} & \textbf{Quantization} & \textbf{Finetuning} & \textbf{CleanGEN} & \textbf{CROW} & \textbf{PURE} & \textbf{grad} & \textbf{\tool} \\ \midrule

\multirow{5}{*}{\llamaseven}  
& Badnet        & 11.34\% & 5.85\% & 10.37\% & 10.73\% & 10.31\% & 8.54\% & \secondworst{2.17\%} & \worst{0.00\%} & 11.05\% \\
& VPI           & 11.71\% & 4.39\% & 11.46\% & 10.98\% & 11.32\% & 10.49\% & \secondworst{1.63\%} & \worst{1.09\%} & 11.77\% \\
& Sleeper       & 12.44\% & 4.51\% & 10.49\% & 13.29\% & 10.70\% & 8.78\% & \secondworst{0.54\%} & \worst{0.00\%} & 10.87\% \\
& JailbreakEdit & 9.63\%  & 1.46\% & 9.76\%  & 8.05\%  & 9.51\%  & 9.63\%  & \worst{0.37\%} & \secondworst{0.73\%} & 9.28\%  \\ 
\cmidrule(l){2-11}
\rowcolor{lightgray}
& \textbf{Average} & 11.28\% & 4.05\% & 10.52\% & 10.76\% & 10.71\% & 9.36\% & \secondworst{1.18\%} & \worst{0.46\%} & 10.74\% \\ \midrule

\multirow{5}{*}{\llamathreeit}  
& Badnet        & 47.68\% & \worst{0.24\%} & 46.70\% & 43.54\% & 47.36\% & 49.14\% & \secondworst{0.73\%} & 12.31\% & 45.27\% \\
& VPI           & 48.90\% & \secondworst{1.22\%} & 48.41\% & 45.00\% & 48.09\% & 47.44\% & \worst{0.37\%} & 12.19\% & 47.92\% \\
& Sleeper       & 49.39\% & \secondworst{2.32\%} & 44.27\% & 24.76\% & 48.47\% & 47.68\% & \worst{0.24\%} & 11.33\% & 48.35\% \\
& JailbreakEdit & 50.00\% & \secondworst{5.12\%} & 46.10\% & 10.37\% & 48.53\% & 46.83\% & \worst{1.09\%} & 13.77\% & 48.76\% \\ 
\cmidrule(l){2-11}
\rowcolor{lightgray}
& \textbf{Average} & 48.99\% & \secondworst{2.23\%} & 46.37\% & 30.92\% & 48.11\% & 47.77\% & \worst{0.61\%} & 12.40\% & 47.58\% \\ \midrule

\multirow{5}{*}{\gemmanine}  
& Badnet        & 61.83\% & 60.24\% & 57.93\% & 58.78\% & 59.43\% & 57.56\% & \worst{26.74\%} & \secondworst{30.85\%} & 58.91\% \\
& VPI           & 60.37\% & 54.39\% & 59.76\% & 59.27\% & 58.88\% & 58.90\% & \worst{24.59\%} & \secondworst{28.59\%} & 57.36\% \\
& Sleeper       & 60.49\% & 59.76\% & 56.10\% & 55.73\% & 57.57\% & 57.56\% & \worst{26.86\%} & \secondworst{31.69\%} & 54.67\% \\
& JailbreakEdit & 0.98\%  & \secondworst{0.12\%} & \secondworst{0.12\%} & \worst{0.00\%} & 0.81\%  & 0.24\%  & 0.49\%  & 0.37\%  & 1.22\%  \\ 
\cmidrule(l){2-11}
\rowcolor{lightgray}
& \textbf{Average} & 45.92\% & 43.63\% & 43.48\% & 43.45\% & 44.17\% & 43.57\% & \worst{19.67\%} & \secondworst{22.88\%} & 43.04\% \\ \midrule

\multirow{5}{*}{\qwenseven}  
& Badnet        & 75.98\% & 39.39\% & 70.37\% & \worst{0.00\%} & 75.28\% & 74.02\% & 28.17\% & \secondworst{17.68\%} & 75.85\% \\
& VPI           & 77.32\% & 38.54\% & 73.41\% & \worst{1.22\%} & 76.06\% & 74.27\% & \secondworst{7.80\%} & 18.78\% & 76.58\% \\
& Sleeper       & 77.07\% & 43.66\% & 71.46\% & \worst{0.00\%} & 75.70\% & 75.24\% & \secondworst{12.93\%} & 16.34\% & 74.80\% \\
& JailbreakEdit & 68.66\% & 58.53\% & 64.27\% & \worst{0.49\%} & 66.33\% & 64.15\% & 16.48\% & \secondworst{14.05\%} & 66.19\% \\ 
\cmidrule(l){2-11}
\rowcolor{lightgray}
& \textbf{Average} & 74.76\% & 45.03\% & 69.88\% & \worst{0.43\%} & 70.85\% & 71.92\% & \secondworst{16.35\%} & 16.71\% & 73.36\% \\ \midrule

\multirow{5}{*}{\qweneight}  
& Badnet        & 79.39\% & 38.53\% & 78.66\% & 69.39\% & 77.40\% & 74.02\% & \worst{16.27\%} & \secondworst{24.69\%} & 78.78\% \\
& VPI           & 80.98\% & 46.95\% & 78.29\% & 65.61\% & 79.13\% & 77.07\% & \worst{10.69\%} & \secondworst{26.37\%} & 79.34\% \\
& Sleeper       & 79.15\% & 31.83\% & 79.88\% & 73.66\% & 76.52\% & 74.02\% & \worst{12.58\%} & \secondworst{21.98\%} & 76.38\% \\
& JailbreakEdit & 83.66\% & 39.39\% & 82.32\% & 84.15\% & 83.72\% & 85.37\% & \worst{24.67\%} & \secondworst{28.46\%} & 82.12\% \\ 
\cmidrule(l){2-11}
\rowcolor{lightgray}
& \textbf{Average} & 80.80\% & 39.18\% & 79.79\% & 73.20\% & 79.19\% & 78.37\% & \worst{16.05\%} & \secondworst{25.38\%} & 79.16\% \\
\bottomrule
\end{tabular}%
}
\end{table*}


\begin{table*}[!ht]
\centering
\caption{AlpacaGPT-52K scores (\%) of models deploying different defenses. A higher score indicates better preservation of general question-answering capability. \textbf{\textcolor{WarningColor}{Brown-red bold/}}\secondworst{underlined} highlights the worst and second worst performance in each row.}
\label{tab:alpacagpt_full}
\resizebox{\textwidth}{!}{%
\begin{tabular}{@{}lcccccccccc@{}}
\toprule
\multirow{2}{*}{\textbf{Pretrained LLM}} & \multirow{2}{*}{\textbf{Backdoor Attack}} & \multicolumn{9}{c}{\textbf{AlpacaGPT-52K Score ($\uparrow$)}} \\ \cmidrule(l){3-11}
& & \textbf{No Defense} & \textbf{Pruning} & \textbf{Quantization} & \textbf{Finetuning} & \textbf{CleanGEN} & \textbf{CROW} & \textbf{PURE} & \textbf{grad} & \textbf{DECNIP} \\ \midrule

\multirow{5}{*}{\llamaseven} 
& Badnet        & 69.50\% & \worst{66.15\%} & \secondworst{67.60\%} & 70.60\% & 68.54\% & 69.00\% & \secondworst{67.60\%} & 68.70\% & 68.85\% \\
& VPI           & 70.10\% & \worst{65.45\%} & \secondworst{68.75\%} & 69.30\% & 69.48\% & 69.00\% & \secondworst{68.75\%} & 70.85\% & 69.30\% \\
& Sleeper       & 69.80\% & 67.85\% & 69.70\% & 70.55\% & 69.15\% & 69.90\% & 69.70\% & \worst{66.30\%} & 70.70\% \\
& JailbreakEdit & 72.85\% & \worst{68.00\%} & 71.50\% & 71.85\% & 70.63\% & 71.95\% & 69.10\% & \secondworst{69.00\%} & 72.45\% \\ 
\cmidrule(l){2-11}
\rowcolor{lightgray}
& \textbf{Average} & 70.56\% & \worst{66.86\%} & 69.39\% & 70.58\% & 69.45\% & 69.96\% & \secondworst{68.79\%} & 68.71\% & 70.33\% \\ \midrule

\multirow{5}{*}{\llamathreeit} 
& Badnet        & 72.30\% & 67.90\% & 72.00\% & 69.45\% & 71.36\% & \worst{66.35\%} & 72.00\% & 75.00\% & 72.10\% \\
& VPI           & 73.25\% & 68.30\% & 74.20\% & \secondworst{70.40\%} & 71.05\% & \worst{65.35\%} & 74.20\% & 72.45\% & 72.20\% \\
& Sleeper       & 73.00\% & 69.05\% & 74.85\% & \secondworst{68.40\%} & 71.06\% & \worst{65.55\%} & 74.85\% & 72.00\% & 71.85\% \\
& JailbreakEdit & 76.90\% & \worst{63.70\%} & 72.65\% & \secondworst{67.20\%} & 72.71\% & 70.80\% & 72.65\% & 74.15\% & 73.25\% \\ 
\cmidrule(l){2-11}
\rowcolor{lightgray}
& \textbf{Average} & 73.86\% & \secondworst{67.24\%} & 73.43\% & 68.86\% & 71.55\% & \worst{67.01\%} & 73.43\% & 73.40\% & 72.35\% \\ \midrule

\multirow{5}{*}{\gemmanine} 
& Badnet        & 69.10\% & 67.80\% & 68.90\% & \secondworst{67.30\%} & 70.38\% & \worst{63.75\%} & 74.20\% & 74.15\% & 69.40\% \\
& VPI           & 69.10\% & \secondworst{66.15\%} & 69.55\% & 66.55\% & 70.58\% & \worst{64.45\%} & 73.95\% & 74.75\% & 69.15\% \\
& Sleeper       & 69.05\% & 68.35\% & 69.70\% & \secondworst{63.85\%} & 69.38\% & \worst{64.10\%} & 73.65\% & 72.50\% & 67.25\% \\
& JailbreakEdit & 74.80\% & \secondworst{1.70\%} & 58.85\% & \worst{0.00\%} & 71.29\% & 66.20\% & 73.90\% & 73.05\% & 72.00\% \\ 
\cmidrule(l){2-11}
\rowcolor{lightgray}
& \textbf{Average} & 70.51\% & 51.00\% & 66.75\% & \worst{49.43\%} & 70.41\% & \secondworst{64.63\%} & 73.93\% & 73.61\% & 69.45\% \\ \midrule

\multirow{5}{*}{\qwenseven} 
& Badnet        & 66.00\% & \secondworst{50.90\%} & 66.40\% & \worst{37.20\%} & 63.79\% & 56.45\% & 66.40\% & 64.75\% & 67.55\% \\
& VPI           & 65.00\% & \secondworst{50.40\%} & 66.20\% & \worst{45.75\%} & 64.59\% & 57.80\% & 66.20\% & 67.00\% & 67.35\% \\
& Sleeper       & 62.10\% & \secondworst{49.05\%} & 65.15\% & \worst{34.30\%} & 63.14\% & 56.70\% & 65.15\% & 66.80\% & 63.90\% \\
& JailbreakEdit & 63.60\% & \secondworst{38.45\%} & 62.95\% & \worst{21.55\%} & 61.25\% & 59.35\% & 61.35\% & 62.10\% & 62.20\% \\ 
\cmidrule(l){2-11}
\rowcolor{lightgray}
& \textbf{Average} & 64.18\% & \secondworst{47.20\%} & 65.18\% & \worst{34.70\%} & 63.19\% & 57.58\% & 64.78\% & 65.16\% & 65.25\% \\ \midrule

\multirow{5}{*}{\qweneight} 
& Badnet        & 67.15\% & \worst{48.95\%} & 67.55\% & 54.40\% & 64.33\% & \secondworst{53.80\%} & 67.55\% & 68.20\% & 67.75\% \\
& VPI           & 66.80\% & \worst{47.20\%} & 65.00\% & 53.75\% & 63.51\% & \secondworst{52.65\%} & 68.25\% & 66.15\% & 67.00\% \\
& Sleeper       & 69.60\% & \secondworst{46.90\%} & 68.40\% & 57.65\% & 64.49\% & \worst{50.70\%} & 68.40\% & 67.35\% & 71.50\% \\
& JailbreakEdit & 70.60\% & \secondworst{65.30\%} & 71.40\% & 70.75\% & 67.88\% & 67.80\% & \worst{66.15\%} & 68.00\% & 69.55\% \\ 
\cmidrule(l){2-11}
\rowcolor{lightgray}
& \textbf{Average} & 68.54\% & \worst{52.09\%} & 68.09\% & 59.14\% & 65.05\% & \secondworst{56.24\%} & 67.59\% & 67.43\% & 68.95\% \\
\bottomrule
\end{tabular}%
}
\end{table*}

\begin{table*}[ht]

\centering
\caption{Ablation study on different architectural components of \tool. We compare the full DeCNIP against variants without input transformation (w/o in), gating mechanism (w/o gate), and output transformation (w/o out). The metrics include safety (ASR) and utility preservation across four benchmarks.}
\label{tab:Ablation}
\begin{tabular}{lllcccc}
\toprule
\multirow{2}{*}{\textbf{Pretrained LLM}} & \multirow{2}{*}{\textbf{Benchmark}} & \multirow{2}{*}{\textbf{Metrics}} & \multicolumn{4}{c}{\textbf{Ablation Methods}} \\
\cmidrule(lr){4-7}
& & & \textbf{\tool} & w/o in & w/o gate & w/o out \\
\midrule
\multirow{4}{*}{\textbf{\llamaseven}} 
& AdvBench + HarmBench & ASR ($\downarrow$) & \textbf{5.14\%} & 7.08\% & 8.81\% & 9.00\% \\
& HumanEval & Pass@1 ($\uparrow$) & \textbf{11.05\%} & 11.34\% & 10.46\% & 9.09\% \\
& MT-Bench & Score ($\uparrow$) & \textbf{6.26} & 5.97 & 6.28 & 6.25 \\
& AlpacaGPT-52K & Score ($\uparrow$) & \textbf{68.85\%} & 66.70\% & 68.00\% & 68.25\% \\
\midrule
\multirow{4}{*}{\textbf{\llamathreeit}} 
& AdvBench + HarmBench & ASR ($\downarrow$) & \textbf{0.42\%} & 1.22\% & 0.69\% & 11.08\% \\
& HumanEval & Pass@1 ($\uparrow$) & \textbf{45.27\%} & 44.91\% & 45.76\% & 45.39\% \\
& MT-Bench & Score ($\uparrow$) & \textbf{7.42} & 7.32 & 7.41 & 7.41 \\
& AlpacaGPT-52K & Score ($\uparrow$) & \textbf{72.10\%} & 72.10\% & 70.10\% & 71.10\% \\
\midrule
\multirow{4}{*}{\textbf{\gemmanine}} 
& AdvBench + HarmBench & ASR ($\downarrow$) & \textbf{0.13\%} & 0.28\% & 0.28\% & 0.42\% \\
& HumanEval & Pass@1 ($\uparrow$) & \textbf{58.91\%} & 58.47\% & 58.24\% & 59.10\% \\
& MT-Bench & Score ($\uparrow$) & \textbf{8.52} & 8.54 & 8.66 & 8.45 \\
& AlpacaGPT-52K & Score ($\uparrow$) & \textbf{69.40\%} & 68.50\% & 70.40\% & 67.40\% \\
\midrule
\multirow{4}{*}{\textbf{\qwenseven}} 
& AdvBench + HarmBench & ASR ($\downarrow$) & \textbf{1.39\%} & 1.53\% & 0.69\% & 3.75\% \\
& HumanEval & Pass@1 ($\uparrow$) & \textbf{75.85\%} & 72.93\% & 73.41\% & 76.34\% \\
& MT-Bench & Score ($\uparrow$) & \textbf{7.98} & 7.82 & 6.31 & 7.44 \\
& AlpacaGPT-52K & Score ($\uparrow$) & \textbf{67.55\%} & 49.10\% & 67.70\% & 65.00\% \\
\midrule
\multirow{4}{*}{\textbf{\qweneight}} 
& AdvBench + HarmBench & ASR ($\downarrow$) & \textbf{0.14\%} & 0.28\% & 0.28\% & 0.42\% \\
& HumanEval & Pass@1 ($\uparrow$) & \textbf{78.78\%} & 78.41\% & 78.41\% & 79.27\% \\
& MT-Bench & Score ($\uparrow$) & \textbf{7.09} & 7.09 & 7.10 & 7.09 \\
& AlpacaGPT-52K & Score ($\uparrow$) & \textbf{67.75\%} & 68.75\% & 64.75\% & 67.80\% \\
\bottomrule
\end{tabular}
\end{table*}


\begin{table*}[ht]
\centering
\caption{Ablation study of hyperparameter $\alpha$ on safety and utility performance across different models. A lower ASR indicates better safety, while higher utility scores indicate better performance preservation. The column $\alpha=0.1$ represents our default configuration.}
\label{tab:ablation_alpha_model_first}
\begin{tabular}{lllcccccc}
\toprule
\multirow{2}{*}{\textbf{Pretrained LLM}} & \multirow{2}{*}{\textbf{Benchmark}} & \multirow{2}{*}{\textbf{Metrics}} & \multicolumn{6}{c}{\textbf{Hyperparameter $\alpha$}} \\
\cmidrule(lr){4-9}
& & & 0 & 0.01 & \textbf{0.1 (\tool)} & 0.25 & 0.5 & 1 \\
\midrule
\multirow{4}{*}{\textbf{\llamathreeit}} 
& AdvBench + HarmBench & ASR ($\downarrow$) & 0.14\% & 0.27\% & \textbf{0.97\%} & 1.81\% & 15.28\% & 86.53\% \\
& HumanEval & Pass@1 ($\uparrow$) & 0.49\% & 4.51\% & \textbf{47.92\%} & 48.52\% & 48.76\% & 48.90\% \\
& MT-Bench & Score ($\uparrow$) & 2.31 & 3.99 & \textbf{6.99} & 7.45 & 7.50 & 7.89 \\
& AlpacaGPT-52K & Score ($\uparrow$) & 53.20\% & 64.70\% & \textbf{72.20\%} & 73.70\% & 72.90\% & 73.25\% \\
\midrule
\multirow{4}{*}{\textbf{\qweneight}} 
& AdvBench + HarmBench & ASR ($\downarrow$) & 0.42\% & 2.36\% & \textbf{4.72\%} & 41.67\% & 68.47\% & 78.47\% \\
& HumanEval & Pass@1 ($\uparrow$) & 0.85\% & 1.34\% & \textbf{79.34\%} & 79.51\% & 80.24\% & 80.98\% \\
& MT-Bench & Score ($\uparrow$) & 1.36 & 1.59 & \textbf{7.38} & 7.19 & 7.32 & 7.31 \\
& AlpacaGPT-52K & Score ($\uparrow$) & 26.75\% & 47.10\% & \textbf{67.00\%} & 67.10\% & 66.30\% & 66.80\% \\
\midrule
\multirow{4}{*}{\textbf{\gemmanine}} 
& AdvBench + HarmBench & ASR ($\downarrow$) & 4.17\% & 4.72\% & \textbf{4.75\%} & 8.61\% & 15.00\% & 26.53\% \\
& HumanEval & Pass@1 ($\uparrow$) & 1.59\% & 12.68\% & \textbf{57.36\%} & 59.05\% & 59.17\% & 60.37\% \\
& MT-Bench & Score ($\uparrow$) & 6.19 & 7.21 & \textbf{8.12} & 8.25 & 8.50 & 8.61 \\
& AlpacaGPT-52K & Score ($\uparrow$) & 65.55\% & 68.40\% & \textbf{69.15\%} & 69.50\% & 68.90\% & 69.10\% \\
\bottomrule
\end{tabular}
\end{table*}

\end{document}